\documentclass[fleqn,usenatbib]{mnras}

\usepackage{newtxtext,newtxmath}
\usepackage[T1]{fontenc}
\usepackage{ae,aecompl}

\usepackage{graphicx}	% Including figure files
\usepackage{amsmath}	% Advanced maths commands
\usepackage{amssymb}	% Extra maths symbols

\newcommand{\teff}{\ensuremath{T_{\mathrm{eff}}}}
\newcommand{\feh}{\ensuremath{[\mathrm{Fe/H}]}}

\newcommand{\logg}{\mbox{$\log g$}}
\newcommand{\jkzero}{$(J-K)_{0}$}
\newcommand{\clr}{$(J-K)_{0}$}

\newcommand{\glx}{\textsc{galaxia}}
\newcommand{\glh}{{\sl GALAH}}

\newcommand{\ktwohermes}{{\sl K2-HERMES}}

\newcommand{\tesshermes}{{\sl TESS-HERMES}}
\newcommand{\apg}{{\sl APOGEE}}

\newcommand{\rave}{{\sl RAVE}}
\newcommand{\lamost}{{\sl LAMOST}}
\newcommand{\gaia}{{\sl Gaia}}
\newcommand{\gcs}{{\sl GCS}}
\newcommand{\twomass}{{\sl 2MASS}}

\newcommand{\galahapogee}{{\sl GADR14RC}}
\newcommand{\apgrcold}{{\sl ADR12RC}}
\newcommand{\apgrcnew}{{\sl ADR14RC}}

\newcommand{\vlos}{$V_{\mathrm{los}}$}

\newcommand{\vphi}{$V_{\mathrm{\phi}}$}
\newcommand{\mnvR}{$\overline{V}_{\mathrm{R}}$}
\newcommand{\mnvz}{$\overline{V}_{\mathrm{z}}$}
\newcommand{\mnvphi}{$\overline{V}_{\mathrm{\phi}}$}

\newcommand{\kms}{{\rm km s}$^{-1}$}
\title[Velocity fluctuations in the Milky Way]{The GALAH Survey: Velocity fluctuations in the Milky Way using Red Clump giants}

\author[Shourya Khanna et al.]{
Shourya Khanna,$^{1,2}$\thanks{E-mail: skha2680@uni.sydney.edu.au}
Sanjib Sharma,$^{1,2}$
Joss Bland-Hawthorn,$^{1,2,3}$
Michael Hayden,$^{1,2}$
\newauthor David M. Nataf,$^{4}$
Yuan-Sen Ting,$^{5,6,7}$
Janez Kos,$^{1,2}$
Sarah Martell,$^{8}$
Toma\v{z} Zwitter,$^{9}$
\newauthor Gayandhi De Silva,$^{1,10}$
Martin Asplund,$^{11}$
Sven Buder,$^{12}$
Ly Duong,$^{11}$
Jane Lin,$^{11}$
\newauthor Jeffrey D. Simpson,$^{10}$
Borja Anguiano,$^{13}$
Jonathan Horner,$^{14}$
Prajwal R. Kafle,$^{15}$
\newauthor Geraint F. Lewis,$^{1}$
Thomas Nordlander,$^{2,11}$
Rosemary F.G. Wyse,$^{4}$
\newauthor Robert A. Wittenmyer,$^{16,17}$
and Daniel B. Zucker$^{1,2}$
\\
$^{1}$Sydney Institute for Astronomy, School of Physics, A28, The University of Sydney, NSW, 2006, Australia\\
$^{2}$Centre of Excellence for Astrophysics in Three Dimensions (ASTRO-3D), Australia\\
$^{3}$Miller Professor, Miller Institute, UC Berkeley, Berkeley CA 94720\\
$^{4}$Center for Astrophysical Sciences and Department of Physics and Astronomy, The Johns Hopkins University, Baltimore, MD 21218, USA\\
$^{5}$Institute for Advanced, Princeton, NJ 08540, USA\\
$^{6}$Department of Astrophysical Sciences, Princeton University, Princeton, NJ 08544, USA\\
$^{7}$Observatories  of  the  Carnegie  Institution  of  Washington,  813  Santa Barbara Street, Pasadena, CA 91101, US\\
$^{8}$School of Physics, University of New South Wales, Sydney, NSW 2052, Australia\\
$^{9}$Faculty of Mathematics and Physics, University of Ljubljana,
Jadranska 19, 1000 Ljubljana, Slovenia\\
$^{10}$Australian Astronomical Observatory, 105 Delhi Rd, North Ryde, NSW 2113, Australia\\
$^{11}$Research School of Astronomy \& Astrophysics, Australian National University, ACT 2611, Australia\\
$^{12}$Max Planck Institute for Astronomy (MPIA), Koenigstuhl 17,
69117 Heidelberg, Germany\\
$^{13}$Department of Astronomy, University of Virginia,P.O. Box 400325, Charlottesville, VA 22904-4325, USA\\
$^{14}$University of Southern Queensland, Toowoomba, Qld 4350, Australia\\
$^{15}$ICRAR, The University of Western Australia,35 Stirling Highway, Crawley, WA 6009, Australia\\
$^{16}$University of Southern Queensland, Computational Engineering and Science Research Centre, Toowoomba, Queensland 4350, Australia\\
$^{17}$Australian Centre for Astrobiology, University of New South Wales, Sydney, NSW 2052, Australia
}

\date{Accepted XXX. Received YYY; in original form ZZZ}

\pubyear{2018}

\begin{document}
\label{firstpage}
\pagerange{\pageref{firstpage}--\pageref{lastpage}}
\maketitle

% Abstract of the paper
\begin{abstract}
If the Galaxy is axisymmetric and in dynamical equilibrium, we expect negligible fluctuations in the residual line-of-sight velocity field. Recent results using the \apg{} survey find significant fluctuations in velocity for stars in the midplane ($|z|<$0.25 kpc) out to 5 kpc, suggesting that the dynamical influence of non-axisymmetric features i.e., the Milky Way's bar, spiral arms and merger events extends out to the Solar neighborhood. Their measured power spectrum has a characteristic amplitude of 11 \kms{} on a scale of 2.5 kpc. The existence of such large-scale streaming motions has important implications for determining the Sun's motion about the Galactic Centre.
Using Red Clump stars from \glh{} and \apg{}, we map the line-of-sight velocities around the Sun (d$<$5 kpc), and $|z|<$1.25 kpc from the midplane. By subtracting a smooth axisymmetric model for the velocity field, we study the residual fluctuations and compare our findings with mock survey generated by \glx{}. 
We find negligible large-scale fluctuations away from the plane. In the mid-plane, we reproduce the earlier \apg{} power spectrum but with 20\% smaller amplitude (9.3 \kms{}) after taking into account a few systematics (e.g., volume completeness). Using a flexible axisymmetric model the power-amplitude is further reduced to 6.3 \kms{}.
Additionally, our simulations show that, in the plane, distances are underestimated for high-mass Red Clump stars which can lead to spurious power-amplitude of about 5.2 \kms{}. 
Taking this into account, we estimate the amplitude of real fluctuations to be $<$4.6 \kms{}, about a factor of three less than the \apg{} result. 
\end{abstract}

% Select between one and six entries from the list of approved keywords.
% Don't make up new ones.

%%%%%%%%%%%%%%%%%%%%%%%%%%%%%%%%%%%%%%%%%%%%%%%%%%

%%%%%%%%%%%%%%%%% BODY OF PAPER %%%%%%%%%%%%%%%%%%

\section{Introduction}

The Milky Way is a large late-type disk galaxy. While the Galaxy has had a quiescent accretion history and is not thought to have experienced a major merger in the past 10 Gyr \citep{2008ApJ...683..597S,2016ARA&A..54..529B}, it is orbited by nearby dwarf galaxies, some of which can cross the disk and perturb it. Some of these orbiting galaxies become disrupted by these encounters, torn asunder to create streams of material orbiting the galaxy such as the Sagittarius stream \citep{2003ApJ...599.1082M}. In addition, the Milky Way also hosts a central bar extending out to 5 kpc \citep{2015MNRAS.450.4050W}, the dynamical effect of which can also be seen in the Solar neighborhood as structures in velocity space. Prominent examples of such structures include the Hercules stream  \citep[e.g.,][]{1998AJ....115.2384D,2010ApJ...725.1676B,2018MNRAS.474...95H}. While it would seem natural to assume kinematic non-axisymmetry at small radii it is worth investigating whether the dynamical effects of the bar extend out to large radii such as around the Solar neighbourhood. Studying the velocity substructure in the Milky Way thus holds clues to large scale evolutionary processes in the Galaxy.

Over the last two decades local surveys such as \gcs{} \citep[Geneva-Copenhagen Survey,][]{2004A&A...418..989N} and \rave{} \citep[Radial Velocity Experiment,][]{2013MNRAS.436..101W} have mapped the Solar neighbourhood extensively and shown evidence of velocity gradients in the disk. With the advent of large-scale surveys it is now possible to venture out of the Solar neighbourhood. For example, with \apg{} \citep[Apache Point Observatory Galactic Evolution Experiment,][]{2016AN....337..863M} the Galactic disk in the mid-plane ($|z|<0.25$ kpc) has been mapped out to 15 kpc. The synergy with other ongoing surveys such as \glh{} \citep[GALactic Archaeology with HERMES,][]{2017MNRAS.465.3203M} and \lamost{} \citep[Large sky Area Multi-Object fiber Spectroscopic Telescope,][]{2012arXiv1206.3569Z}, now allows us to study the region away from the mid-plane and attempt to visualise kinematics and chemistry in 3D.

The limitations of small scale surveys can be understood in the context of the Local Standard of Rest (LSR)\footnote{The local standard of rest (LSR) is defined as the frame of reference of a star at the Sun's location that is on a circular orbit in the Galactic gravitational potential. It is thus assumed that the LSR has no radial or vertical motion w.r.t to the Galactic centre, as suggested by the proper motion of Sgr A* which shows that such motion is negligible \citep{2010MNRAS.403.1829S}.} which is generally defined based on results from the \gcs{} survey to be $(U,V,W)_{\odot}=(11.1,12.24,7.25)$ \kms{} \citep{2010MNRAS.403.1829S}. However, it has been suggested that the `true' LSR may differ from this current standard, most notably through the study of kinematics of \apg{} Red Clump stars by \citep{2015ApJ...800...83B} (hereafter B15). In their analysis they subtract an axisymmetric model for the line-of-sight velocity field and find significant residual bulk motion or streaming of about 11 \kms{}. Fourier analysis of the residual motion shows that the scale of fluctuations is about 3 kpc i.e., much larger than the Solar neighbourhood. B15 suggest that the whole solar neighbourhood is moving with respect to the Galaxy on a non-axisymmetric orbit due to perturbations from the Galactic Bar. Furthermore, taking this large scale streaming motion into account, they suggest that the value of V$_{\odot}$, the Sun's motion relative to the circular velocity ($V_{\mathrm{circ}}$), be revised upwards to 22.5 \kms{}, implying that the Solar neighbourhood is moving ahead of the LSR. Although, the proper motion of Sgr A* is well constrained at $6.379 \pm 0.024$ mas/yr \citep{2004ApJ...616..872R}, there is still uncertainty on the distance of the Sun from the Galactic center (R$_{\mathrm{\odot}}$). It is thus important to have multiple methods to constrain the Solar motion about the Galactic center. \citep{2017arXiv170406274R} take this forward by making use of highly accurate proper motions from \gaia{} data release 1 \citep{2016A&A...595A...2G} and \rave{}-DR4 to model the asymmetric drift and leave the Solar motion as a free parameter. They propose a new, much lower value for V$_{\odot}$ of 0.94 \kms{}, but again their analysis is only restricted to the Solar vicinity probing well within 2 kpc of the Sun with mean distances around 1 kpc. 

Using a purely astrometric sample from \gaia-TGAS, \citep{2017A&A...602L..13A} detect velocity asymmetries of about 10 \kms{} between positive and negative Galactic longitudes. Once again, however, their study is only based on proper motions of the stars involved. The detected asymmetry seems directed away from the Solar neighbourhood and towards the outer disk. Similarly, using SDSS-DR12 white dwarf kinematics, \cite{2017MNRAS.469.2102A} find $\partial V_{\rm R} /\partial R =-3\pm 5$ \kms{} and an asymmetry in $<V_{\rm z}>$ between the population above and below the plane of the Galaxy. 

Given the compelling evidence of non-equilibrium kinematics shown by a diverse stellar population, it would clearly be interesting to perform a large-scale 3D analysis of the Milky Way. In this paper we examine the line-of-sight kinematics of Red Clump (RC) stars selected from the \glh{} and \apg{} spectroscopic surveys. In general, observed data has a non-trivial selection function, and in some cases leaves a strong signature on the data. Not taking this into account can lead to spurious fluctuations in the velocity distribution of the target stars. Hence it is imperative to check and compare the results to those obtained through use of a synthetic catalog of stars. To this end we make use of axisymmetric galaxy models using the \glx{}\footnote{\glx{} is a stellar population synthesis code based on the Besancon Galactic model by \cite{2003A&A...409..523R}. \glx{} uses its own 3D extinction scheme to specify the dust distribution and the isochrones to predict the stellar properties are from the Padova database \citep{2008A&A...482..883M,1994A&AS..106..275B}. Full documentation is available at \href{http://galaxia.sourceforge.net/Galaxia3pub.html}{http://galaxia.sourceforge.net/Galaxia3pub.html}} code {\citep{2011ApJ...730....3S}. 

Throughout the paper we adopt a right handed coordinate frame in which the Sun is at $R_{\odot}$=8.0 kpc from the Galactic center and has Galactocentric coordinates ($X_{\rm gc}$,$Y_{\rm gc}$,$Z_{\rm gc}$) = (-8.0,0,0) kpc. 
The cylindrical coordinate angle $\phi$ increases in the anti-clockwise direction. The rotation of the Galaxy is clockwise in the ($X_{\rm gc},Y_{\rm gc}$) plane.  The heliocentric Cartesian frame is related to Galactocentric by $X_{\rm hc}=X_{\rm GC}+8$, $Y_{\rm hc}=Y_{\rm GC}$ and $Z_{\rm hc}=Z_{\rm GC}$. 
$X_{\rm hc}$ is negative toward $\ell=180^\circ$ and $Y_{\rm hc}$ is positive towards Galactic rotation. For transforming velocities between heliocentric and Galactocentric frames we use $(X_{\rm gc,\odot},Y_{\rm gc,\odot},Z_{\rm gc,\odot})=(U_{\odot},\Omega_{\odot}R_{\odot},W_{\odot})$.
Following \cite{2010MNRAS.403.1829S}, we adopt $(U,W)_{\odot}=(11.1,7.25)$ \kms{}, while for the azimuthal component we use the constraint of  $\Omega_{\odot}=30.24$ \kms{}kpc$^{-1}$ which is set by the proper motion of Sgr A*, i.e., the Sun's angular velocity around the Galactic center \citep{2004ApJ...616..872R}.

The structure of the paper is as follows: In \autoref{sec:red_clump_sel_brief} we briefly describe our Red Clump selection scheme, which includes using \glx{} to calibrate de-reddened colors against spectroscopic parameters in order to select a pure Red Clump sample. This is used to derive distances in \autoref{sec:distances}. Then in \autoref{sec:data} we briefly discuss the observed and simulated datasets used in the paper. Our kinematic model and methods are described in \autoref{sec:kinematic_model}. In \autoref{sec:high_mass_stars} we test our model on a \glx{} all-sky sample and identify high mass Red Clump population as a contaminant. Next, in \autoref{sec:mid-plane_data} we analyse observed data in the midplane and compare with the \apg{} result of B15. Then in \autoref{sec:Data_vs_Galaxia} we extend the analysis to the offplane region and compare our results with selection function matched \glx{} realizations before discussing our findings in \autoref{sec:discussion}.

%###########################################
% Red Clump SELECTION & CALIBRATION (BRIEF)
%###########################################
\section{Selecting pure Red Clump sample to estimate distances}
\label{sec:red_clump_and_dist} 
\subsection{Red Clump calibration and selection}
\label{sec:red_clump_sel_brief} 
The Red Clump (RC) is a clustering of red giants on the Hertzsprung-Russell diagram (HRD), that have gone through helium flash and now are quietly fusing helium in the convective core. These stars on the helium-burning branch of the HRD, have long been considered `standard candles' for stellar distances as they have very similar core masses and luminosities \citep{1970MNRAS.150..111C}. While many studies use the RC, there is considerable variation in the literature over calibration for the absolute magnitude of these stars. Studies of RC stars using parallaxes from Hipparcos have shown that their average absolute magnitude in the $K_{s}$ (hereafter $K$) band spans the range -1.65$<M_{Ks}<$-1.50 \citep{2016ARA&A..54...95G} and there is ongoing effort to revise this using \gaia{} \citep{2017arXiv170508988H}. The color dependence of the Red Clump on population parameters is also well known, for example the $(J-Ks)$ color is predicted to increase by 0.046 mag from ([Fe/H],[$\alpha$/Fe])=(-0.30,+0.10) to (0.00,0.00) \citep{2016MNRAS.456.2692N}. Given this variation, in this work we choose not to assume single $M_{Ks}$ value to estimate distances but instead derive an empirical relation between $K$ band absolute magnitude and metallicity [Fe/H]. 
We choose the $K$ band for two reasons. While some passbands are more affected than others by metallicity variations within the RC population, such effects seem to be greatly reduced in the $K$ band \citep{2002MNRAS.337..332S}. This, combined with the fact that the $K$ band is least affected by extinction, makes it a reliable passband in which to derive fundamental properties of the RC population.

However, we will first need to select a reliable sample of Red Clump stars.Our selection function is based in terms of de-reddened colors $C_{JK}=$\jkzero{} and the spectroscopic parameters: surface gravity \logg, metallicity [Fe/H] and effective temperature \teff{} as described in \cite{2014ApJ...790..127B}. In the \apg{} Red Clump catalog \citep{2014ApJ...790..127B} the photometry is corrected for extinction using the Rayleigh Jeans Color Excess method \citep[RJCE;][]{2011ApJ...739...25M} which requires photometry in 2MASS and [4.5 $\mu m$] bands. However, it is difficult to get de-reddened colors accurately from photometry alone. 
So, to overcome this, we use pure Red Clump stars from \glx{} to derive empirical relations expressing $C_{JK}$ in terms of [Fe/H] and \teff{}. This allows us to derive de-reddened colors from spectroscopic parameters, which we can then use to select the Red Clump samples for any given spectroscopic sample. In particular, the \glx{} Red Clump sample is also used to obtain the aforementioned $M_{Ks}$-[Fe/H] curve (see \autoref{tab:mks_feh}), which is used to estimate distances. The procedure above is described in full detail in \autoref{app:red_clump_sel}. 

%====================
% TABLE: Contamination
\begin{table*}
	\centering
	\caption{Accuracy of the Red Clump selection function as predicted by all-sky $J<15$ mock samples from \glx{}. Shown are results for cases with: 1) no uncertainty on spectroscopic parameters; 2) with uncertainty typically expected from high resolution spectra, e.g., \apg{} \citep{2015AJ....150..148H}    and 3) same as 2) but with smaller $\sigma_{\log T_{\rm eff}}$ which should be achievable with good quality spectra.}
	\label{tab:contamination}
	\begin{tabular}{l|lll|lll} 
    \hline
    & $\sigma_{\log T_{\rm eff}}$ & $\sigma_{\rm [Fe/H]}$ & $\sigma_{\log g}$ 
    & Recall\footnote{We define Red Clump as those stars that satisfy equations (\ref{logg_cut}-\ref{clr_cut}). So, recall refers to fraction of total number of Red Clump stars that are selected.} 
    & Precision\footnote{Here precision refers to fraction of selected stars that have initial stellar mass $>M_{\rm RGB, tip}$} & $\sigma_{\rm dmod}$\footnote{The quantity in brackets denotes $\sigma_{\rm dmod}$ for the actual Red Clump stars, which is not significantly affected by addition of spectroscopic uncertainties.} \\
    & dex   & dex   & dex           & (\%)   & (\%) & mag\\
    \hline
    1  & 0.0      & 0.0   & 0.0 & 100.0 & 97.9 & 0.10 (0.09)  \\ 
    2  & 0.011    & 0.05  & 0.1 & 77.0  & 93.6 & 0.16 (0.12)\\ 
    3  & 0.0055   & 0.05  & 0.1 & 87.0 & 94.8 & 0.12 (0.11)\\ 
    \hline
	\end{tabular}
\end{table*}
%====================

We now check the accuracy of our selection function in recovering the Red Clump stars. For this we compute precision (fraction of selected stars that are part of the Red Clump) and recall (fraction of Red Clump stars that are selected), which 
are two commonly used measures of accuracy in the field of information retrieval.
We find that 97.6\% of our selected stars are part of the Red Clump.  
Since our selection is based on spectroscopic parameters that have uncertainties associated with them, we also explored the effects of adding Gaussian errors of $(\sigma_{\log T_{\rm eff}}, \sigma_{\rm [Fe/H]}, \sigma_{\log g})= (0.011,0.05,0.1)$ dex. For $T_{\rm eff}=4700$ K, the typical temperature of a Red Clump star, the uncertainty in temperature is $120$ K. In spite of the uncertainties,  the precision of the selected stars was found to be 83\% (see \autoref{tab:contamination} for summary), however, the recall 
dropped to 69\%. If the uncertainty on temperature is reduced by a factor of two, the precision increases to 91\% and recall to 85\%, suggesting that it is important get precise and accurate temperatures.

%%%%%%%%%%%%%%%%%%%%%%%%%
\subsection{Distances} 
\label{sec:distances}
We now proceed to estimate distances for our Red Clump sample. The distance modulus for a given passband $\lambda$ corrected for extinction is given by
\begin{equation}
\label{dmod}
d_{\rm mod}= m_{\lambda}-M_{\lambda}-A_{\lambda},
\end{equation} with apparent magnitude $m_{\lambda}$, absolute magnitude $M_{\lambda}$ and extinction $A_{\lambda}$. For the $K$ band, $M_{\lambda}$ is derived using our $M_{K}$-[Fe/H] relation (\autoref{tab:mks_feh}), while for the extinction we make use of the derived intrinsic colors $C_{JK}({\rm [Fe/H]},T_{\rm eff})$\footnote{ 
See equation \ref{direct_calib_eqn}} i.e.,
\begin{equation}
(J-K)  -  C_{JK}= (A_{J}-A_{K}),
\end{equation}
and this can be related to the reddening E(B-V) using $f_{\lambda}=A_{\lambda})/E(B-V)$ as
\begin{equation}
(A_{J}-A_{K}) = (f_{J}-f_{K}) E(B-V).
\end{equation}
After rearranging, we get the general relation  
\begin{equation}
\label{extinction}
A_{\lambda} = f_{\lambda} \times \frac{(J-K) - C_{JK}}{f_{J} - f_{K}},
\end{equation} 
with $f_{\lambda}$ as in \cite{1998ApJ...500..525S}
(see \autoref{tab:tab_extinct}).

To illustrate the accuracy achieved in estimating distances for the \glx{} sample, we show the residuals in $d_{\mathrm{mod}}$ with respect to the true distance modulus  in \autoref{fig:dist_accuracy}. The residuals lie close to zero, with a  typical distance uncertainty of 4\%. There is also no significant bias as a function of metallicity [Fe/H]. 
If uncertainty in spectroscopic parameters are taken into 
account the dispersion in estimated distance modulus $\sigma_{\rm dmod}$ increases and 
this is shown in \autoref{tab:contamination} for some typical cases. The main reason for the increase in $\sigma_{\rm dmod}$ is the 
contamination from stars that are not Red Clump, e.g., RGB stars, which can be understood from the quoted precisions in the table. 
The quantity in brackets 
denotes $\sigma_{\rm dmod}$ for the actual Red Clump stars, which is not significantly affected 
by addition of spectroscopic uncertainties. 

 %============================
% FIGURE: distance accuracy (allsky)
\graphicspath{{figures/}}  
\begin{figure}
\includegraphics[width=\columnwidth]{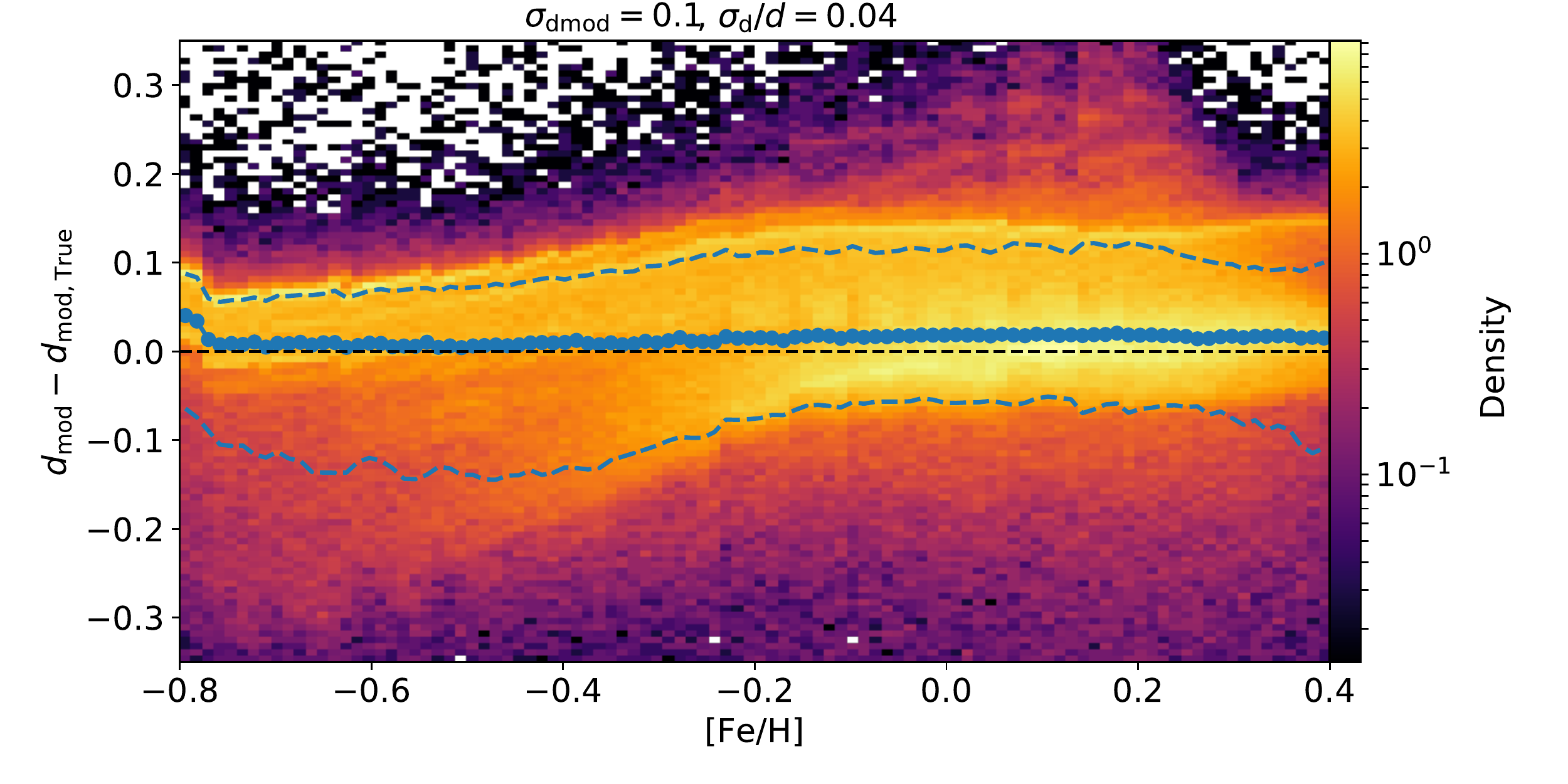}
\caption{{Distance accuracy using Red Clump calibration on \glx: The residuals in the distance modulus are concentrated close to zero (black dotted line) and there is no significant bias with metallicity. The blue dotted lines indicate the 1$\sigma$ bounds, with the overall distance error being 4 \%.}{\label{fig:dist_accuracy}}}
\end{figure}
%====================

%%%%%%%%%%%%%%%%%%%%%%%%%%%
%============================
\section{Data and methods} 
\label{data_and_methods}
%============================ 
% FIGURE: Data coverage 
\graphicspath{{figures/}} 
\begin{figure*}
\centering          
\includegraphics[width=\columnwidth]{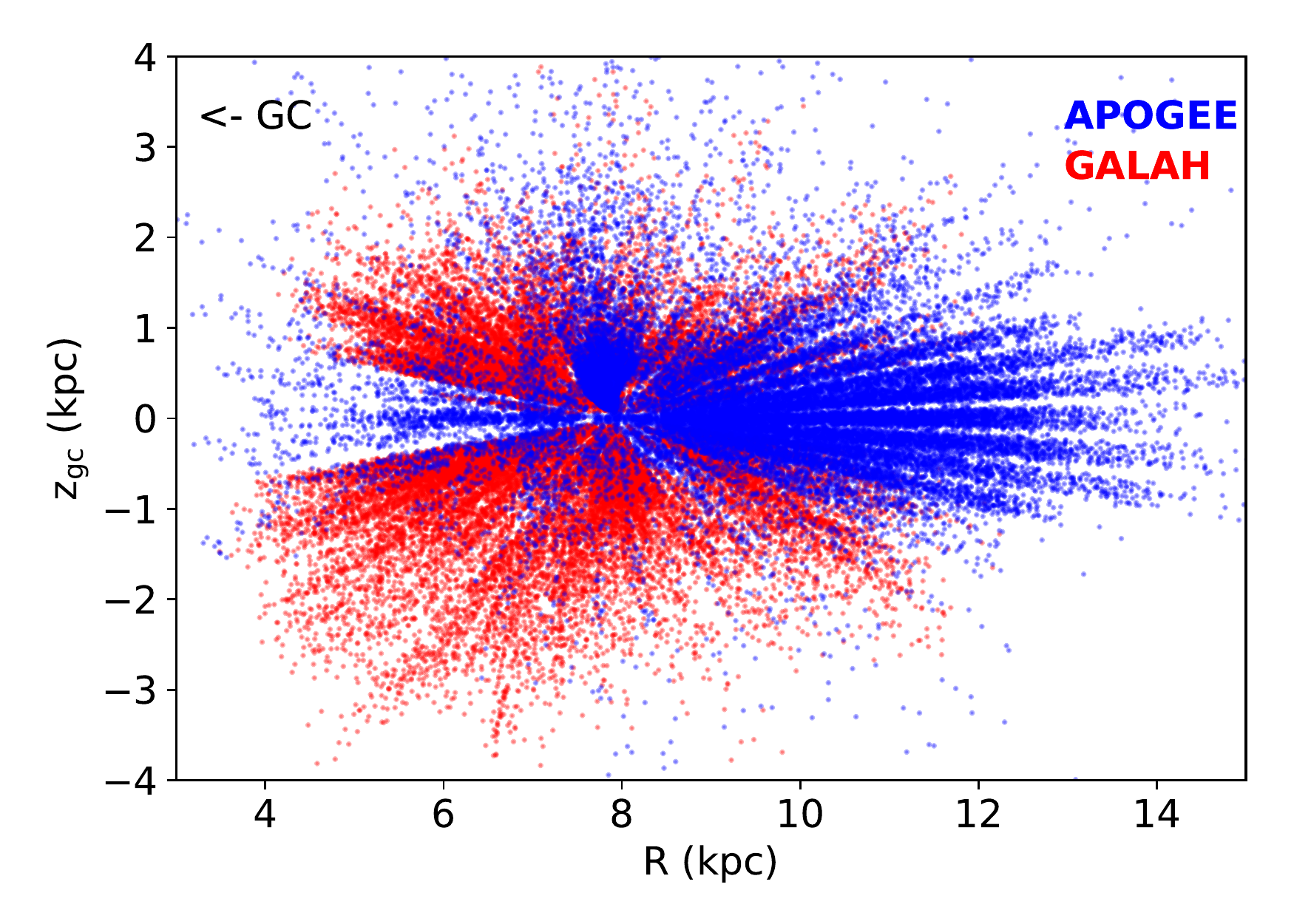}
\caption{{Distribution of the combined Red Clump dataset GADR14RC in Galactocentric $R-z$ plane. While the \apg{} coverage dominates in plane and towards the anti-center, \glh{} surveys the off-plane region more extensively.}{\label{fig:datacov_Rz}}}
\includegraphics[width=\columnwidth]{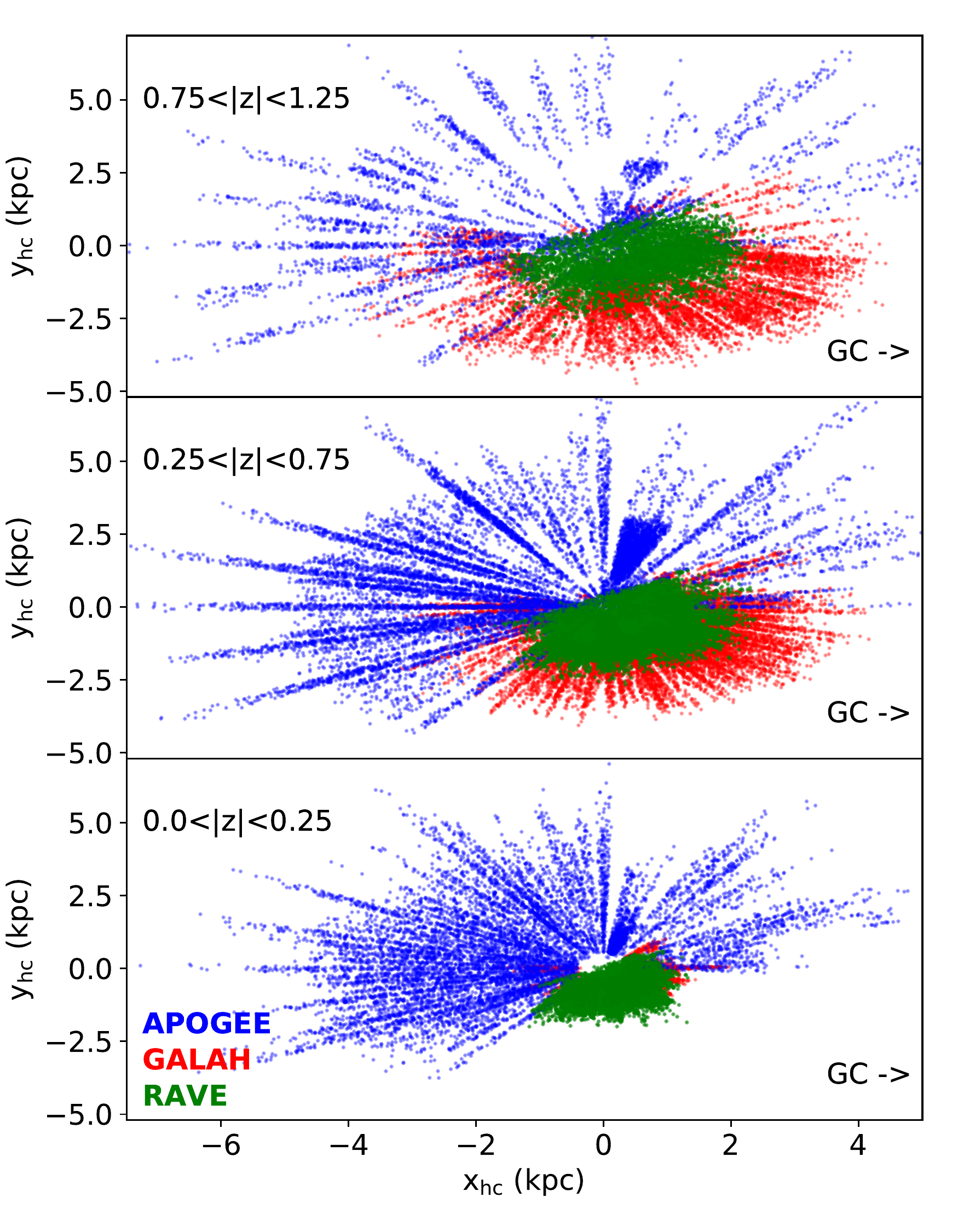}
\caption{{Distribution of GADR14RC in heliocentric $(x,y)$ coordinates where the symmetrical regions above and below the plane have been merged together in slices in $z$ (kpc). \apg{} probes deep into the disk while \glh{} provides good coverage moving away from the plane, and to illustrate this also shown is the coverage of RAVE DR5. 
}{\label{fig:datacov_xy_zslices}}}
\end{figure*}
%============================ 

\subsection{Datasets}
\label{sec:data}
In this paper we make use of data from the \apg{} and \glh{} surveys from which Red Clump stars are selected using the selection scheme described in \autoref{app:red_clump_sel} unless otherwise specified. Following is a brief overview of the datasets used for our analysis:

We downloaded the Red Clump catalog\footnote{\href{https://data.sdss.org/sas/dr12/apogee/vac/apogee-rc/cat/}{APOGEE DR12-RC fits files}} of \apg{} DR12 \citep{2014ApJ...790..127B}, in order to compare our results directly with B15. This dataset contains 19937 stars and will be referred to as \apgrcold{}. Similarly we also obtained the latest available RC catalog\footnote{\href{https://data.sdss.org/sas/dr14/apogee/vac/apogee-rc/cat/}{APOGEE DR14-RC fits files}} from \apg{} DR14 \citep{2018ApJS..235...42A}. This contains 29502 stars and will be referred to as \apgrcnew{}. In both cases, while we do not apply our Red Clump selection method, we do estimate the distances using the scheme in \autoref{sec:distances}. Our distances 
were found to be in excellent agreement with those in the \apg{} Red Clump catalog. In our analysis the distances are used to compute velocity maps, and we found that there was no difference between the velocity maps computed using either of the distances.

Where it appears the additional tag `SF\_Bovy' explicitly means that the dataset used has exact selection as in the \apg{} Red Clump catalogs.

Next, from the internal release of \glh{} data up to October 2017 we preselect stars in the magnitude range $9<V<14$. 
The data includes fields observed as part of the \ktwohermes{} \citep{2018AJ....155...84W} and \tesshermes{} \citep{2018MNRAS.473.2004S} programs but not the fields observed as part of the pilot\footnote{Data collected before March 2014 i.e., with cob id$<1403010000$ is excluded, where cob id = date*10000 + run no.} survey. Also, data without a proper selection function  (field id <-1) was excluded from the analysis. 
The spectroscopic parameters are from the same pipeline that was used in 
\citet{2018MNRAS.473.2004S} and further details of 
spectroscopic analysis techniques used can be found there and in \citet{2018MNRAS.476.5216D}. 
Details on reduction and estimation of radial velocity  are in \citet{2017MNRAS.464.1259K}.
From the full data we select Red Clump stars using our scheme in \autoref{app:red_clump_sel} and obtain 33183 RC stars. This is merged with \apgrcnew{} to form a combined observed dataset called \galahapogee{} and again where it appears, the additional tag `SF\_New' signifies that our selection method was employed. This combined set provides a more complete $(x,y)$ spatial coverage as shown in \autoref{fig:datacov_Rz} and \autoref{fig:datacov_xy_zslices}, where only for comparison we show 44166 Red Clump stars from \rave{}-DR5 \citep{2017AJ....153...75K} using our selection scheme. The combined dataset allows us to explore the region well beyond the Solar neighbourhood.

To examine the validity of this analysis, we use \glx{} to simulate the selection functions of \apg{}\footnote{\href{https://data.sdss.org/sas/dr14/apogee/target/}{APOGEE DR14 fields}} \citep{2013AJ....146...81Z} and \glh{} \citep{2017MNRAS.465.3203M}, and generate a combined Red Clump dataset using our selection schemes for direct comparison with \galahapogee{}. Finally, for \autoref{sec:high_mass_stars} we also generate an all-sky mock Red Clump catalog to test our kinematical models. All \glx{} samples were generated with the `warp' option turned off in order to allow easier interpretation of our experiments.

%%%%%%%%%%%%%%%%%%%%%
%----------------------------------------------------------------------
\subsection{Proper motions}
\label{sec:proper_motion}
In order to transform from the heliocentric to Galactocentric frame we require highly accurate proper motions. Gaia DR1 has provided high precision parallaxes for about 2 million objects and the DR2 (expected April 2018) will extend this to nearly a billion objects and will also provide proper motions. In the meantime the two extensively used proper motion catalogues PPMXL \citep{2010AJ....139.2440R} and UCAC4 \citep{2013AJ....145...44Z} have been improved using Gaia DR1 positions to produce UCAC5 \citep{2017AJ....153..166Z} and HSOY \citep[\textbf{H}ot \textbf{S}tuff for \textbf{O}ne \textbf{Y}ear,][]{2017A&A...600L...4A}. Until Gaia DR2 these updated catalogues will provide proper motion with 1-5 mas/yr precision. For all our observed datasets, where available, we use the average of UCAC5 \& HSOY values, and default (UCAC4) proper motions elsewhere. We have checked that this has no impact on our results. Moreover, our main analysis does not 
make use of proper motions. 

%%%%%%%%%%%%%%%%%%%%%%%%%%%%%%%%%%%%%%%%%%%%%%%%%%
% Velocity Fluctuations/ Kinematics
%%%%%%%%%%%%%%%%%%%%%%%%%%%%%%%%%%%%%%%%%%%%%%%%%%
\subsection{Kinematic model}
\label{sec:kinematic_model}
In this section we will describe the framework of our kinematical modelling. Our goal is to reproduce the observed line-of-sight velocity field (\vlos) using an axisymmetric Galactic model. In our scheme the Galactocentric velocity distribution, $V=(V_{R},V_{z},V_{\phi})$, follows the triaxial Gaussian distribution
\begin{equation}
p(V|r,\tau) = \frac{1}{(2\pi^{3/2})\sigma_{R}\sigma_{\phi}\sigma_{z}}
\exp \bigg\{\frac{V_{R}^{2}}{\sigma_{R}^{2}} + \frac{V_{z}^{2}}{\sigma_{z}^{2}} + \frac{(V_{\phi} -\overline{V}_{\phi} )^{2}}{\sigma_{\phi}^{2}}  \bigg\} ,
\end{equation}
where we assume that \mnvR{} and \mnvz{} are negligible. The mean Galactocentric azimuthal velocity \mnvphi{} can be written  using \cite{1946ApJ...104...12S} as:
%============================ 
% EQUATION: mean Vphi eqn 
\begin{eqnarray}
\label{mn_vphi_BT1} 
\overline{V}_{\phi}^{2} = V_{\rm circ}^{2}(R,z) + \overbrace{\sigma_{R}^{2} \bigg(\frac{\rm{d} \ln \rho}{\rm{d} \ln R} + \frac{\rm{d} \ln \sigma_{R}^{2}}{\rm{d} \ln R}   + 1 - \frac{\sigma_{\phi}^{2}}{\sigma_{R}^{2}} + 1 - \frac{\sigma_{z}^{2}}{\sigma_{R}^{2}} \bigg)}^{V_{asym}},  
\end{eqnarray} 
where $V_{\rm circ}$ is the Galactocentric circular velocity, and $V_{\rm asym}$ is the asymmetric drift. Assuming exponential density profiles for the Galactic disk \citep[$\rho \propto e^{-R/R_{d}}$,][]{2013ApJ...773..183S}  and velocity dispersion $(\sigma \propto e^{\frac{R-R_{\odot}}{R_{\sigma}}})$ we get  
\begin{eqnarray}
\label{mn_vphi_BT} 
\overline{V}_{\phi}^{2} = V_{\rm circ}^{2}(R,z) +\sigma_{R}^{2}\bigg(-\frac{R}{R_{\mathrm{d}}} -\frac{2R}{R_{\sigma}} + 1 - \frac{\sigma_{\phi}^{2}}{\sigma_{R}^{2}} + 1 - \frac{\sigma_{z}^{2}}{\sigma_{R}^{2}} \bigg) 
\end{eqnarray} 
However, equation \ref{mn_vphi_BT} is valid only for the case where the principle axis of the velocity ellipsoid is aligned with the spherical coordinate system $(r,\theta,\phi)$ centered on the Galactic center, i.e, $\overline{V_{R}V_{z}} =  (V_{R}^{2} - V_{z}^{2}) (z/R)$ \citep{2008gady.book.....B}. There is however, evidence to suggest that the ellipsoid is aligned with the cylindrical system $(R,\phi,z)$ \citep[e.g.,][]{2014MNRAS.437..351B}, in which case $\partial \overline{V_{R}V_{z}}/\partial z = 0$ and the term $1 - \frac{\sigma_{z}^{2}}{\sigma_{R}^{2}}$ drops out from equation \ref{mn_vphi_BT}. Since the actual answer probably lies in between the two alignments, we instead take into account the contribution of dispersion terms ($\sigma_{\phi,R,z}$) as a new parameter $c_{\rm ad}$,
%============================
% EQUATION: Vphi Bovy model
%
\begin{equation}
\label{vphi_sig_full}
\overline{V}_{\phi}^{2}=V_{\rm circ}^{2}(R,z) +\sigma_{R}^{2}\bigg(-\frac{R}{R_{\mathrm{d}}} -\frac{2R}{R_{\sigma}} + c_{\rm ad}\bigg).
\end{equation} 
%If $\sigma_{\phi,R,z} \propto \exp(-R/R_{\sigma})  
%f_{\rm AVR}(\tau)$, where $f_{\rm AVR}$ is the age velocity dispersion 
%relation and $\tau$ the age, the $c_{\rm ad}$ still remains a constant. 
%====================
% TABLE: sigma-z interp
\begin{table}
\centering
\caption{Parameters to model the velocity dispersion $\sigma_R$ as a function of height $|z|$.\label{tab:sig_z_rel}}
\begin{tabular}{l|l|l|l|l}
\hline
\hline
${|z|}$    & 0.0  & 0.5 & 1.0 & 2.0 \\
\hline
$\delta_{\sigma_{0}}$ & 0.0  & 5.0 & 7.0 & 7.0 \\	
\hline
\end{tabular}
\end{table}
%====================
Using the above framework we can now describe the individual models employed:
\begin{itemize}
\item \textit{Bovy1}: The model used by B15 is derived in \citet[][B12 hereafter]{2012ApJ...759..131B}. Essentially they assume $\partial \overline{V_{R}V_{z}}/\partial z = 0$, exponential surface density profile, exponential velocity dispersion profile and a constant circular velocity and then use the distribution function from \citet{1999AJ....118.1201D} to model the asymmetric drift. 
\citet{2014ApJ...793...51S} fitted the B12 model to \rave{} data and showed that the B12 model can be approximated by setting $c_{\rm ad}=0.28$ in equation \ref{vphi_sig_full}.
In order to reproduce the results of B15 we adopt this value for $c_{\rm ad}$. Furthermore, in accordance with B15, we set $R_{d}=3$ kpc, $R_{\sigma}=\infty$ kpc, $\sigma_{R}=31.4$, and assume a flat profile for the circular velocity $V_{\rm circ}=220$ \kms{}, $V_{\rm circ}+ V_{\odot}=242.5$ \kms{}. We use \textit{Bovy1} only for the mid-plane ($|z|<0.25$ kpc) as was the case in B15. 

\item \textit{globalRz}: B15 model requires making a number of assumptions, e.g., about the circular velocity profile, the $\sigma_{\phi}/\sigma_{R}$ ratio as well as the $\sigma_{\mathrm R}$ profile. Typically, $\sigma_{\mathrm R}$ in the disk lies around 20-40 \kms{}\citep{2012ApJ...759..131B}, however the vertical variation in dispersion requires proper modelling of the AVR and hence a good handle on stellar ages. Moreover, $V_{\rm circ}$ itself has a non-trivial profile, as for example was found with kinematic analysis of \rave{} where gradient in both radial $(\propto  \alpha_{R}(R-R_{\odot}))$ and vertical directions ($\propto \alpha_{z}|z|^{-1.34}$) was reported \citep{2014ApJ...793...51S}. Furthermore, if we compute $\overline{V}_{\phi}$ using proper motions, we see that the  profiles are not flat in R (\autoref{fig:data_vphi_prof}).

Given that some of the assumptions might not be correct, 
for our analysis we adopt a flexible model for $\overline{V}_{\phi}$, that is a 2$^{nd}$ degree multivariate polynomial in cylindrical Galactocentric coordinates $R$ and $z$, more specifically,
%============================
% EQUATION: Global model
%
\begin{equation}
\label{vphi_model}
\overline{V}_{\phi} = {\sum\limits_{i=0}^{2}}{\sum\limits_{j=0}^{2}}  a_{ij}  (R-R_{\odot})^{i}z^{j} ,
\end{equation}
%============================
The model prediction in Galactocentric coordinartes can be transformed to heliocentric coordinates assuming $(U_{\odot},\Omega_{\odot}R_{\odot}, W_{\odot})$ for the solar motion and fitted to observed the line-of-sight velocity $V_{\mathrm{los,mod}}$. The $\Omega_{\odot}$ is given by the the proper motion of Sgr A*, and hence this approach does not require us to assume a value for $V_{\odot}$ or $V_{\mathrm{circ}}$.  In order to fit for the coefficients $a_{ij}$, we assume that the observed \vlos{} is a Gaussian, $\mathcal{N}(.||{\rm mean,dispersion}$), centered at \vlos$_{\mathrm{,mod}}$ with dispersion $\sigma_{\rm los}=31.4$ \kms{} (similar to B15). This can be summarized as,
%============================
% EQUATION: Likelihood functions
%
\begin{equation}
\label{vlos_prob}
p(V_{\mathrm{los}}|a_{ij},l_{\mathrm{gc}},z_{\mathrm{gc}},R) = \textit{$\mathcal{N}$}(V_{\mathrm{los}}|V_{\mathrm{los, mod}},\sigma_{v}) ,
\end{equation}
%============================
and we call this model \textit{globalRz}. The \textit{MCMC} fitting is carried out using the \textit{bmcmc} package \citep{2017arXiv170601629S}.

\item \textit{Strom\_z}: Finally, we will now describe the model for our \glx{} simulations. While we could just use the \textit{globalRz} model to approximate kinematics in \glx{}, however, flexible models like \textit{globalRz} with many free parameters run the risk of overfitting the data. Hence we devise a more realistic model. Note, our aim here is to generate a simple and realistic null hypothesis case, i.e., a smooth axisymmetric model that has no velocity fluctutaions. The default model in \glx{} is based on the Str{\"o}mberg relation with parameters from the RAVE-GAU kinematic model from \citet[][(S14) their last column of Table 6]{2014ApJ...793...51S}. This model is able to describe the z variation in velocity dispersions ($\approx$ AVR), but it requires stellar ages as input. Since for observed data, ages are not available, instead of using the default model, we modify it take the variation of $\sigma_{R}$ with height $z$ into account. For this we adopt the following form for $\sigma_{R}$,
\begin{equation}
\sigma_{R}(R,z) = (\sigma_{0} + \delta_{\sigma_{0}}(|z|)) \exp(-R/R_{\sigma}) .
\label{equ:galaxia_sigma}
\end{equation}
 and fit for $\sigma_{0}$ and $\delta_{\sigma_{0}}$ 
using a mock \glx{} realization with RAVE-GAU model. We find 
$\sigma_{0}=30.7$ \kms{} and $\delta_{\sigma_{0}}$ for three different values of $|z|$ is given in \autoref{tab:sig_z_rel}. To obtain $\delta_{\sigma_{0}}$ for any arbitrary value of $z$ we use linear interpolation.
For the thick disk we assume a mono-age population (11 Gyr old) and we assume that the thick disc obeys the AVR of the thin disc. 
If this is not done the velocity distribution in the upper slices might deviate strongly from a Gaussian distribution and this will lead to velocity fluctuations and will make the simulation unsuitable for out null hypothesis test.

Finally, we fit equation \ref{vphi_sig_full} to the mock \glx{} realization and find the best match for $c_{\rm ad}=0.77$. The disk and velocity dispersion scale lengths are adopted directly from S14, i.e., $(R_{d},R_{\sigma})=(2.5,13.7)$ kpc. Lastly, in \glx{} we use $V_{\rm circ}+ V_{\odot}=226.84 + 12.1$ \kms{} and the circular velocity profile is from \cite{2011ApJ...730....3S} and is not flat.

\end{itemize}

%%%%%%%%%%%%%%%%%%%%%
\subsection{Fourier analysis of velocity fluctuations}
%============================
% FIGURE: 
\graphicspath{{figures/}} 
\begin{figure}  
\centering
\includegraphics[width=.8\columnwidth]{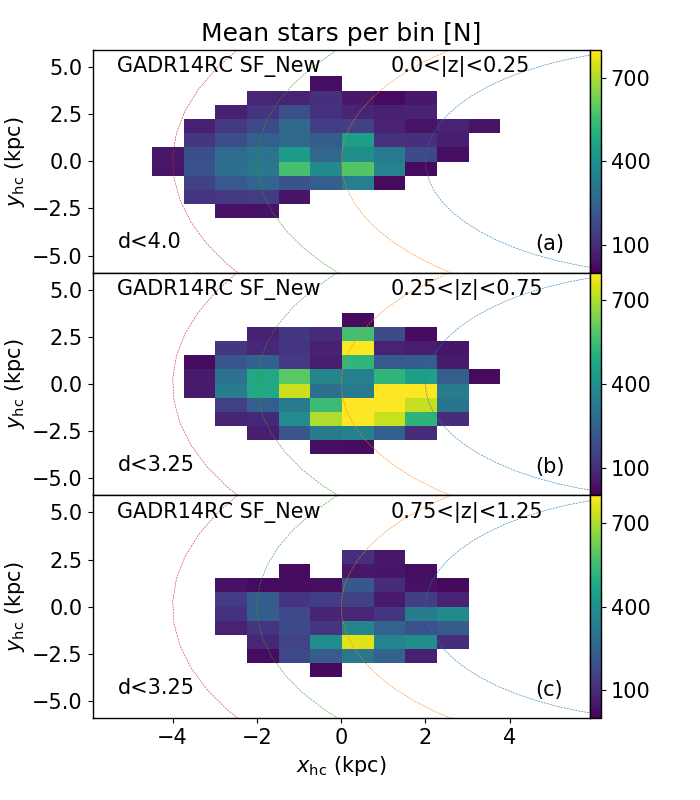}
\caption{Stellar number density of the \galahapogee{} dataset for the three vertical slices used in our analysis. Each pixel has a minimum of 20 stars. The radial cuts applied here correspond to completeness in magnitude limited selection (see \autoref{fig:GADR14_mag} for details).
\label{fig:GADR14RC_num_dens}}
\end{figure}
%============================
Each dataset is divided into three slices in z (as in \autoref{fig:datacov_xy_zslices}), and further binned into $(x,y)$ space with bins of size $0.75\times0.75$ kpc$^{2}$. The resulting stellar density map for each vertical slice of the \galahapogee{} dataset is shown in \autoref{fig:GADR14RC_num_dens}. For each bin, we calculate the residual $\Delta V_{\mathrm{los}}=V_{\mathrm{los}} - V_{\mathrm{los, mod}}$, to produce a 2D velocity fluctuation image $h$. To reduce the contribution from Poisson noise we set $h=0$ for bins than have less than 20 stars.
Next we perform Fourier analysis of the image $h$ and calculate the 2D power spectrum of fluctuations as 
\begin{equation}
P_{kl} = \frac{1}{N_{\rm eff}} |A_{\rm kl}|^{2} \Delta x \Delta y,      
\end{equation}
where $A_{\rm kl}$ is the 2d Fast Fourier Transform (FFT) of the image $h$ and $\Delta x$ and $\Delta y$ are the size of the bins along $x$ and $y$ directions. 
$N_{\rm eff}$ is the effective number of bins in the image and is given by 
$\sum_{i} \sum_{j} H (n_{ij}-20)$, where $H$ is the Heaviside step function 
and $n_{ij}$ is the number of stars in the $(i,j)$-th bin. Next we average $P_{kl}$ azimuthally in bins of $k=\sqrt{k_x^2+k_y^2}$ to obtain the 1D power spectrum $P(k)$. The $P(k)$ as defined above satisfies the following normalization condition given by the \textit{Parseval's} theorem,
\begin{eqnarray}
\int_0^{\infty} \! P(k) 2 \upi k  \, \mathrm{d}k &=& \sum_k \sum_l P_{kl} \Delta k_x \Delta k_y \\ \nonumber
&=& \frac{\sum_{i} \sum_{j} H (n_{ij}-20) h_{ij}^2}{N_{\rm eff}}
\end{eqnarray}
{We present $\sqrt{P(k)}$ that has the dimensions of \kms{} as our final result. 
The presented formalism to compute the power spectrum  is slightly different that of B15, but it matches the results of B15 and importantly ensures that the estimated power spectrum $P(k)$ is invariant to changes in size of the bin, the overall size of the image box, and  bins with missing data.
The noise for the power spectrum is calculated in the same manner except that for the input signal we use normally distributed data with zero mean and dispersion equal to the standard deviation of $\Delta V_{\mathrm{los}}$. }

%============================
% FIGURE: data vphi profile 
\graphicspath{{figures/}} 
\begin{figure}
\centering     
\includegraphics[width=1.0\columnwidth]{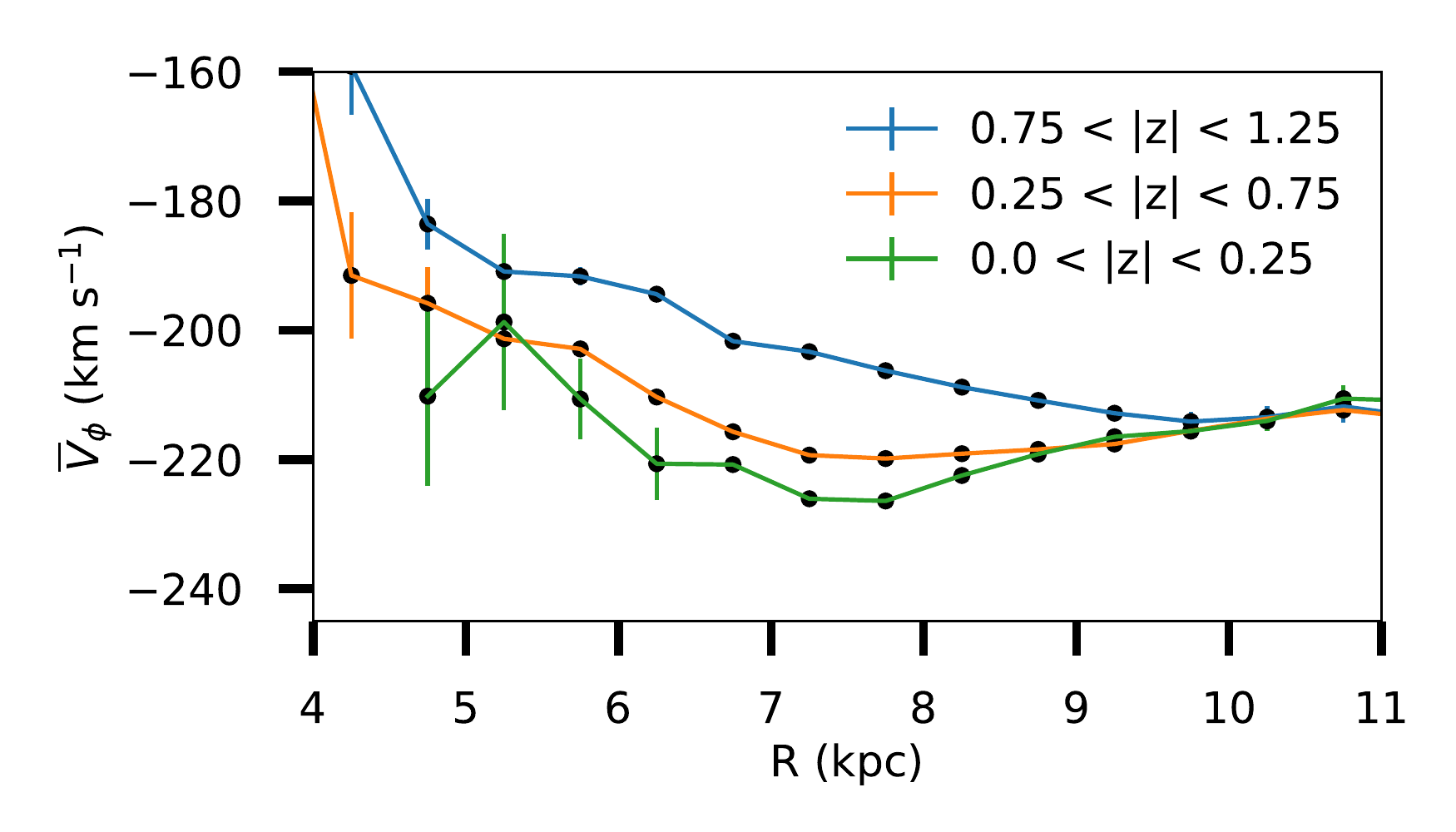}
\caption{{Mean Galactocentric rotation ($\overline{V}_{\phi}$) derived using proper motions for the combined dataset  GADR14RC and shown for different $z$ slices. The profiles look parabolic in nature with a steepening gradient as we move away from the plane.}{\label{fig:data_vphi_prof}}}
\end{figure}
%============================ 

%%%%%%%%%%%%%%%%%%%%%%%%%%%%%%%%%%%%%%%%%%%%%%%%%%
% RESULTS
%%%%%%%%%%%%%%%%%%%%%%%%%%%%%%%%%%%%%%%%%%%%%%%%%%

%============================
% FIGURE: Rogue stars Hmag 
\graphicspath{{figures/}} 
\begin{figure*}
\centering     
\includegraphics[width=2.0\columnwidth]{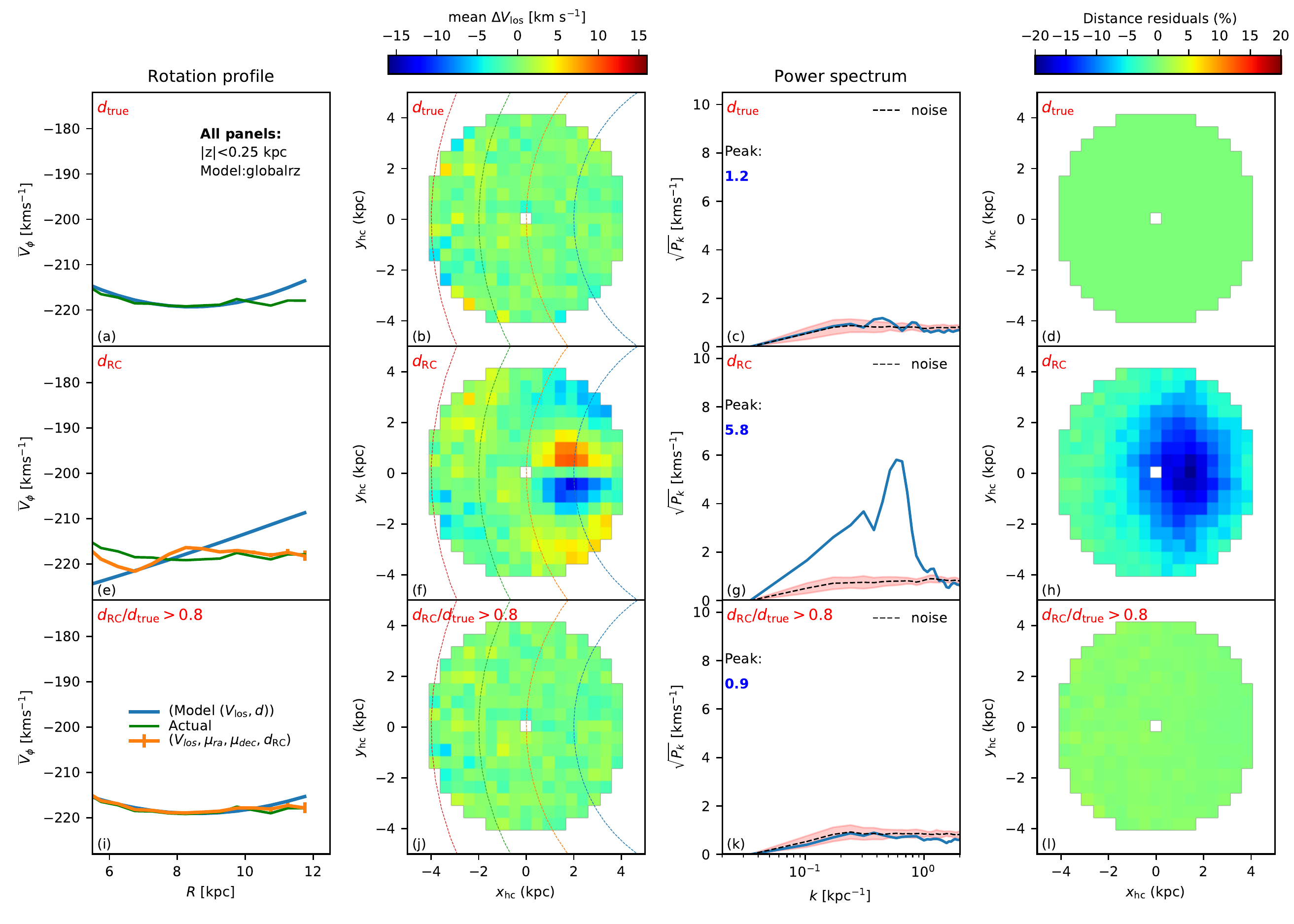}
\caption{Kinematic modelling of \glx{} all-sky sample. 
Results with three different choices of distance are shown,  
true distances (Top panels), Red Clump distances (Middle panels) and Red Clump distances but with stars restricted to $d_{\rm RC}/d_{\rm True}>0.8$ (Bottom panels). Each column shows a different aspect of the kinematics. (a,e,f): The \mnvphi{} as function of $R$ obtained by fitting the \textit{globalRz} model to \vlos{}. The actual \mnvphi{} profile and the profile obtained with 
proper motion and $d=d_{RC}$ is also plotted alongside. 
(b,f,j) The line-of-sight residual velocity map 
obtained after subtracting the best fit \textit{globalRz} model (also overplotted are curves of R = [6,8,10,12] increasing towards negative X$_{hc}$). (c,g,k) Power spectrum of the residual velocity map.
(d,h,l) Map of distance residuals computed with respect 
to $d_{\rm True}$. 
\label{fig:vmap_hmag+rogue_allsky}}
%----------
\end{figure*}

%============================ 
% FIGURE: HMP stars distribution (Mass,vert,d_er vert)
\graphicspath{{figures/}} 
\begin{figure}
\includegraphics[width=1.0\columnwidth]{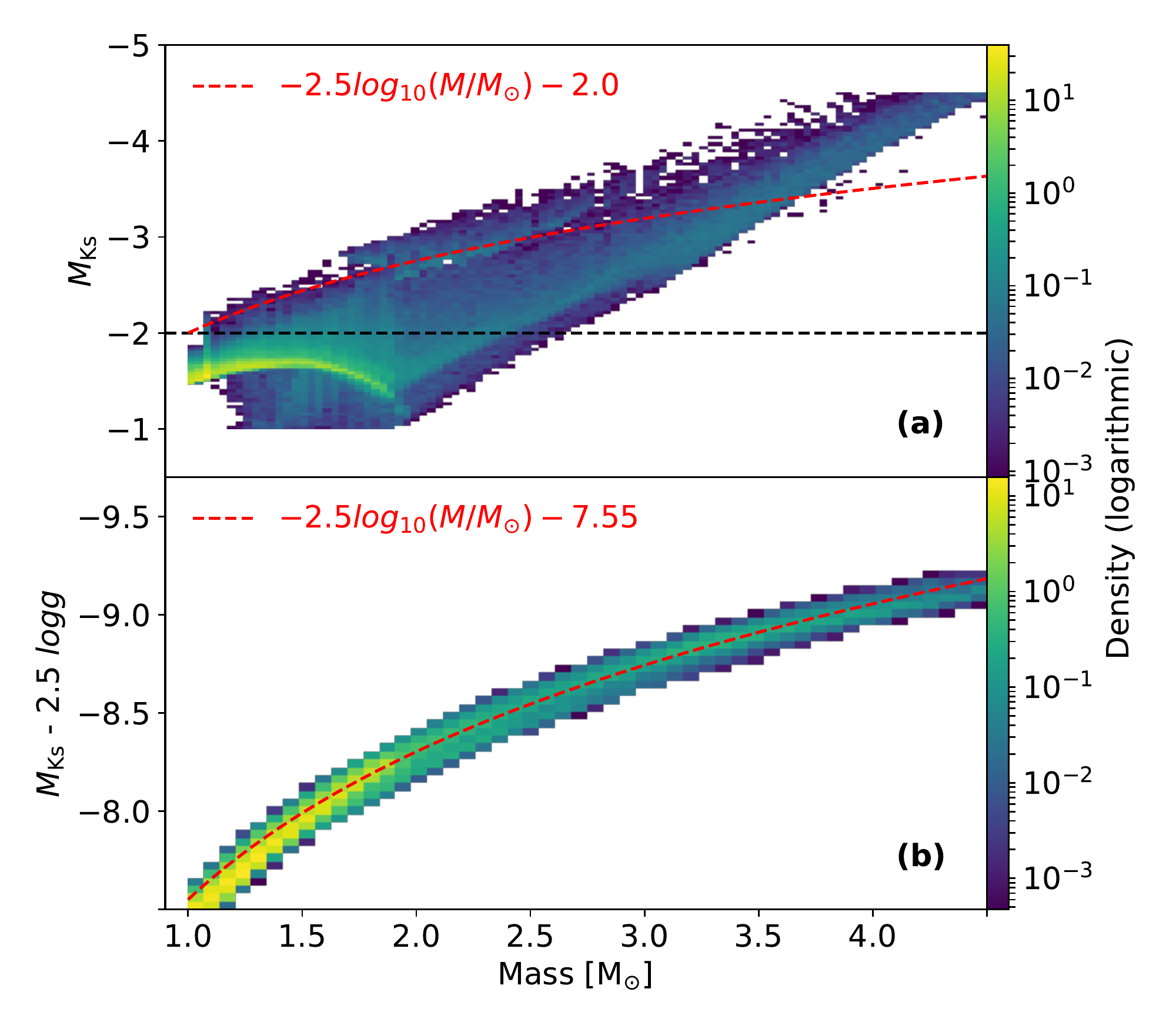}
\caption{Mass distribution of Red Clump stars for a $H<13.8$ sample simulated with {\glx{}}. Panel (a) shows the distribution of Red Clump stars in the $(M_{K_s},{\rm Mass})$ plane. It is clear that the luminous Red Clump stars also have 
higher mass. Stars above the black dashed roughly correspond to where $d_{\rm RC}/d_{\rm True}>0.8$. Panel (b) shows the distribution of Red Clump stars in the $(M_{K_s}-2.5 \log g,{\rm Mass})$ plane. The tight relation is because the Red Clump stars lie in a narrow range of $T_{\rm eff}$. The red dotted line is $= -2.5\log {\rm M}$ in both the panels. 
\label{fig:rogue_allsky_dist2}}
\end{figure}
%============================ 

%============================ 
% FIGURE: HMP stars distribution (Mass,Mks)
\graphicspath{{figures/}} 
\begin{figure}
%rogue_mks
\includegraphics[width=1.0\columnwidth]{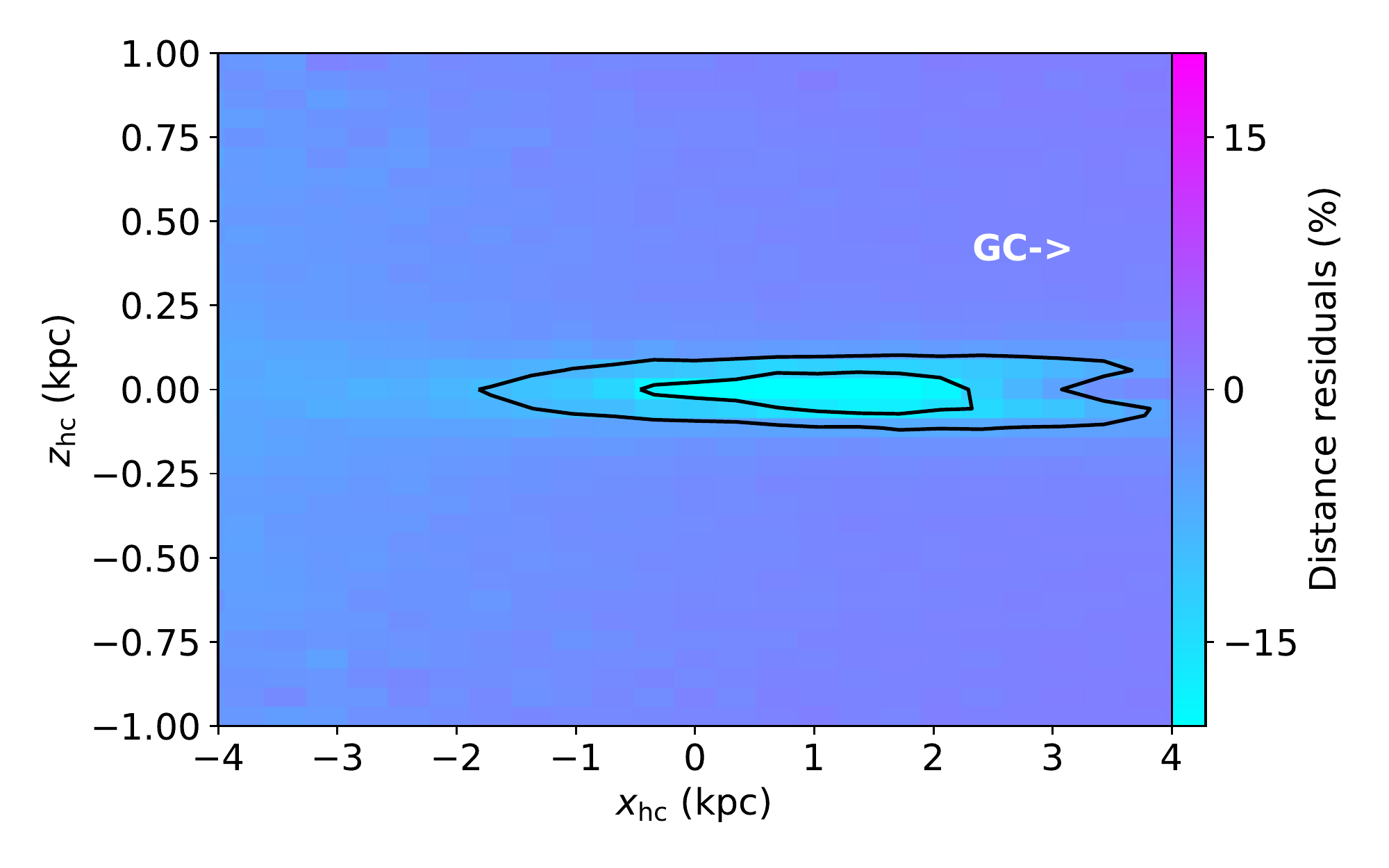}
\caption{Properties of Red Clump stars for a $H<13.8$ sample simulated with \glx{}. We show the map of mean distance residuals $(d_{\rm RC}-d_{\rm True})/d_{\rm True}$. It can be seen that close to the plane and towards the Galactic Center, the distance is underestimated. The contours indicate the  the fraction of Red Clump stars (10\% and 40\% levels shown) that are unusually luminous, identified using $M_{\mathrm{Ks}}<-2.0$. This suggests that distance errors are due to luminous Red Clump stars. \label{fig:rogue_allsky_dist1}}
\end{figure}
%============================ 

\graphicspath{{figures/}}  
\begin{figure}[h]
\includegraphics[width=\columnwidth]{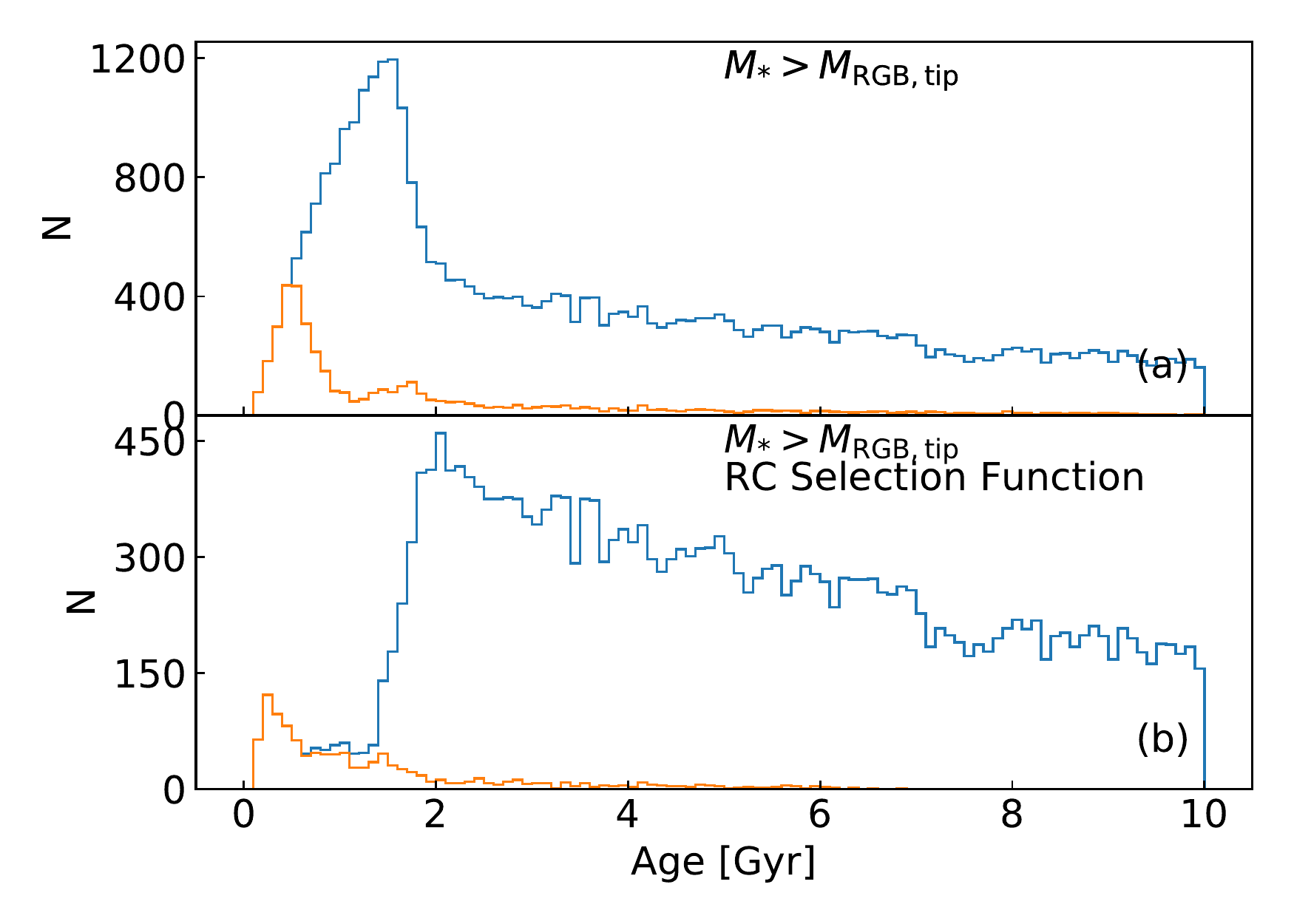}
\caption{Age distribution of \glx{} RC stars with $(1.8< \log g < 3.0)$ and $4300<T_{\rm eff}/{\rm K}<5200$. The RC stars are defined as stars having mass greater than the RGB tipping mass. The star formation is as in \glx{}, which is almost a constant star formation rate. a) The RC sample is selected based on simple cuts in spectroscopic parameters (\logg{},\teff{}) as defined above. The overall age distribution (blue curve) has a peak around 1.5 Gyr and this is dominated by stars with $M_{K}>-2.0$, while the brighter stars (orange curve) with $M_{K}<-2.0$ peak around 0.5 Gyr i.e., are much younger. b) The  RC selection scheme as described in this paper is based on  \citep{2014ApJ...790..127B} (B14) and shown in equations (\ref{logg_cut}-\ref{clr_cut}) is now applied to sample in a). This removes contamination from secondary clump stars (SRC) as well as the RGB bump and in the remaining RC sample, the younger population stands out even more clearly with the majority being Age $<$ 1 Gyr old.\label{fig:age_dist_red_clump}}
\end{figure}

%====================

%============================
% FIGURE: 
\graphicspath{{figures/}} 
\begin{figure*} 
\includegraphics[width=2.0\columnwidth]{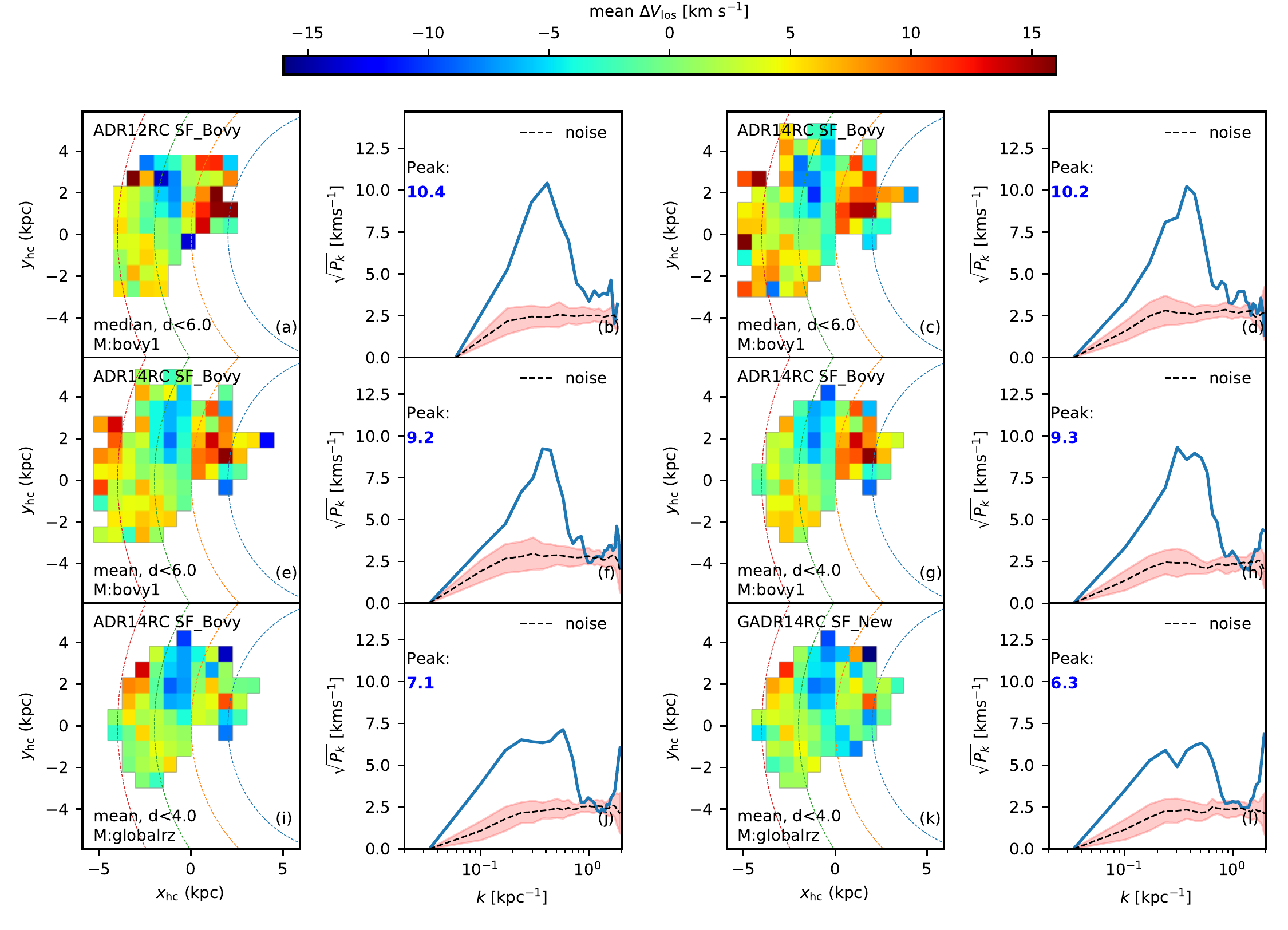} 
      
\caption{Residual velocity maps and power spectrum for the observed data in the $|z|<0.25$ kpc slice. Shown are cases for different data sets, with different radial cuts and kinematic models to illustrate the effect of systematics. (a-b) Data used is \apg{}\_RC\_DR12 with stars restricted to radial distance $d<6$ kpc, using median statistics to compute the residual and using Bovy's analytical model for the kinematics. (c-d) Same as a-b but with \apg{}\_RC\_DR14. (e-f) Same as c-d but now residual is computed using mean statistics. (g-h) Same as e-f but data restricted to $d<4$ kpc to satisfy volume completeness. (i-j) Same as g-h but we now apply the flexible 2d polynomial kinematic model named as \textit{globalRz} to show it reduces power. (k-l) We apply \textit{globalRz} model to the combined \apg{}\_DR14 and \glh{} dataset that uses the new Red Clump selection criteria and distance estimation scheme described in the paper. 
\label{fig:power_spectrum_mid-plane}}
\end{figure*}
%============================

%%%%%%%%%%%%%%%%%%%%%
\section{Results}
\label{sec:results}
The observed data has complicated selection functions in terms of magnitude and spatial coverage. Therefore  before we study the observed data, we will first consider a much simpler dataset using \glx{} that has uniform spatial coverage. This will allow us to test the method described in  \autoref{sec:kinematic_model} and explore any selection function related biases.

\subsection{GALAXIA all-sky sample: High mass Red Clump stars}
\label{sec:high_mass_stars}
%============================
%============================ 
Using \glx{} we generate an all-sky sample that has $H<13.8$, the magnitude boundary of the \apg{} data set 
in the mid-plane, and select Red Clump stars using the scheme in \autoref{app:red_clump_sel}. We make three versions of this dataset, one with true distances ($d=d_{\rm True}$), one with with Red Clump-derived distances ($d=d_{\rm RC}$) and one with Red Clump-derived distances but only for stars with 
$d_{\rm RC}/d_{\rm True} >0.8$. The last of these is chosen to provide a control sample to check for systematic errors in distances.
For each data set we fit the {\it globalRz} kinematic model to \vlos{} data and derive the \mnvphi{} profile and then construct the $\Delta V_{\mathrm{los}}$  map (\autoref{sec:kinematic_model}). In \autoref{fig:vmap_hmag+rogue_allsky}, we only show results for the mid-plane region with $|z|<0.25$ kpc. 
The panels in first column compare the derived \mnvphi{} profile with the actual \mnvphi{} profile, computed directly using line-of-sight motion, proper motions and  true distances. The \mnvphi{} profile computed using Red Clump distance is 
also shown alongside. The panels in second column show the map of velocity fluctuations $\Delta V_{\mathrm{los}}$, while their power spectrum is shown in panels of the third column. The median power spectrum expected due to Poisson noise is shown in dotted black and 68 percentile spread around it
based on 20 random realizations is shown in pink. Finally, in the fourth column we show the map of distance residuals. The results for each case are summarized below.

\begin{itemize}
\item True distances $d=d_{\rm True}$: It is clear that for the true distances we are able to recover the profile by fitting \textit{globalRz} model to \vlos{}. This is also reflected in the  map of $\Delta V_{\mathrm{los}}$, where we obtain a smooth map with negligible residuals. Furthermore, the 1D power spectrum also has amplitude consistent with noise of about 2 \kms. This scenario is as would be expected of a perfectly axisymmetric galaxy.

\item RC distances $d=d_{\rm RC}$: The results are more interesting for the Red Clump derived distances case. Here, the actual \mnvphi{} profile (green line) is not reproduced accurately by the \textit{globalRz} model (blue line) unlike the previous case. The model overestimates the profile beyond the Solar circle ($R_{\odot}$) and underestimates it towards the Galactic center. The \mnvphi{} profile computed using proper motions also does 
not match the actual profile.
The $\Delta V_{\mathrm{los}}$ residual map shows a peculiar dipole along the $y$ axis for $x>0$. This feature gives rise to a sharp peak in the power spectrum with amplitude of \textbf{5.9} \kms at a physical scale of $k^{-1}=1.6$ kpc. Exactly at the location where we see high residuals in \vlos{} we also see high residual in distances.

\item RC distances but only for $d_{\rm RC}/d_{\rm True}>0.8$: The results of this case are very similar to that for case where we use true distances. 
\end{itemize}

For the first case with true distances the residuals in both \vlos{} and distance are zero by definition. For the second case with RC distances, we see significant residuals. It is clear  that the region corresponding to the high $\Delta V_{\mathrm{los}}$ residual also corresponds to high distance residual, i.e., distance errors. This suggests the cause of high residuals is systematic errors in distances. This is further confirmed by the results of the third case, where we restrict the analysis to stars with $d_{\rm RC}/d_{\rm True}>0.8$ and find no residuals in \vlos{} or distances. 

For the second case, the distance residuals are negative which means that the distances are underestimated. This would have the effect of bringing stars closer to us than in reality, more importantly, their kinematics would be inappropriate for their inferred location. This is why we see a dipole in the \vlos{} maps. Since the velocity field is incorrect, the best fit \textit{globalRz} model fails to reproduce the actual \mnvphi{} profile. Due to systematics in distances the velocity profile inferred using proper motion would also be wrong, and this is the reason for the mismatch of the orange line with the green line in \autoref{fig:vmap_hmag+rogue_allsky}e. Again this is confirmed in \autoref{fig:vmap_hmag+rogue_allsky}i, where we restrict stars to $d_{\rm RC}/d_{\rm True}>0.8$ and there is no mismatch between any of the \mnvphi{} profiles.

We now investigate the cause of systematic errors in distances of Red Clump stars. We generate an all sky $H<13.8$ sample with \glx{}, identify Red Clump stars in it, and then study their properties.  \autoref{fig:rogue_allsky_dist2}a, shows the distribution of Red Clump stars in the plane of $M_{\mathrm{K_s}}$ and stellar mass $M$. Typically,  Red Clump stars have $M_{\mathrm{K_s}}\approx-1.60$, however  \autoref{fig:rogue_allsky_dist2}a shows that there is a tail extending down to much brighter magnitudes.
Stars with $d_{RC}/d_{\rm True}<0.8$ that were responsible for strange features in residual velocity maps in \autoref{fig:vmap_hmag+rogue_allsky} correspond to 
$M_{\mathrm{K_s}}<-2$ and this is shown as the black dashed line in the panel. In the tail below the line, brightness is strongly correlated with stellar mass, which extends up to 4 $M_{\odot}$. We know that mass of a red giant star is anti-correlated with age \citep[e.g.,][]{2016AN....337..875S, Miglio2017}, with massive stars being in general younger. So the cause for the systematic errors in the Red Clump distances is the presence 
of  young Red Clump stars that have high mass and luminosity. 

The anti-correlation of absolute magnitude with mass is easy to understand. Red Clump stars lie in a narrow range of $T_{\rm eff}$. Hence their luminosity $L$ is proportional to $R^2$. Given that surface gravity $g=M/R^2$, and since $M_{K}$ represents the Luminosity L well, we have
\begin{eqnarray}
M_{K} \propto -2.5 \log L \propto -2.5 (\log M - \log g) \\
M_{K} -2.5 \log g  \propto -2.5 \log M \label{seismic_rel_two}
\end{eqnarray}
For a given $\log g$, the magnitude decreases with mass 
and the expected trend is shown in \autoref{fig:rogue_allsky_dist2}a. For Red Clump stars $\log g$ is not constant, to take this into account 
in \autoref{fig:rogue_allsky_dist2}b, we  show stars in the 
$(M_K -2.5 \log g, {\rm Mass})$ space. The stars now perfectly follow 
the predicted relation of equation \ref{seismic_rel_two}.

We now investigate as to where we expect to find such high mass stars and in which regions do we expect significant errors in distances.
\autoref{fig:rogue_allsky_dist1} shows the map of distance 
residual in the $(x,z)$ plane. We see that the distance residuals are high in the mid-plane of the Galaxy and towards the Galactic Center. The contours overplotted on \autoref{fig:rogue_allsky_dist1}, 
show the fraction of Red Clump stars that have $M_{\mathrm{K_s}}<-2$, i.e., very luminous. 
Close to the plane and towards the Galactic Center in certain areas the fraction is higher than 0.3. The regions of high distance residuals correspond to region with higher fraction of high-mass Red Clump stars, this provides a causal link for the high distance residuals. 

Why is the contamination from young, high-mass RC so prominent close to the plane and towards the Galactic Center? This is due to a combination of four different effects. Firstly, due to the age scale height relation in the Galaxy, younger stars have smaller scale height and are closer to the plane. Secondly, the  surface density profile of stars in the Galaxy falls off exponentially with distance from the Galactic Center, which means there are more such stars towards the Galactic Center. 
Thirdly, along any given line-of-sight the volume of a cone around it increases as square of the distance. So more stars from far away with larger true distances are displaced to regions with smaller apparent distances. Finally, the spectroscopic selection function designed to select RC stars also plays a role in making the high mass stars appear more prominently.  For constant star formation rate the number of Red Clump stars show a sharp peak around an age of 1.5 Gyr \citep{2016ARA&A..54...95G}. 
But our contaminant bright stars having $M_{Ks}<-2$, peak at 0.5 Gyr 
and are not associated with the peak at 1.5 Gyr. The age distribution 
of RC stars in \glx{} is shown in \autoref{fig:age_dist_red_clump}a, also shown are the 
contaminant bright stars. \autoref{fig:age_dist_red_clump}b shows the age distribution  after applying our RC selection function. The peak at 1.5 Gyr vanishes but not the one at 0.5 Gyr. It is clear  that the selection function introduces a strong age bias rejecting a significant fraction of young stars, but the young contaminant bright stars are not rejected, instead they become more prominent. 

We also studied the off plane slices and found no peculiar features in the residual velocity maps. This is expected as the contamination from high-mass RC stars does not extend far away from the mid-plane. 

%%%%%%%%%%%%%%%%%%%%%
\subsection{Velocity fluctuations in the mid-plane for observed data}
\label{sec:mid-plane_data}
%============================
% FIGURE: 
\graphicspath{{figures/}} 
\begin{figure}  
\includegraphics[width=1.0\columnwidth]{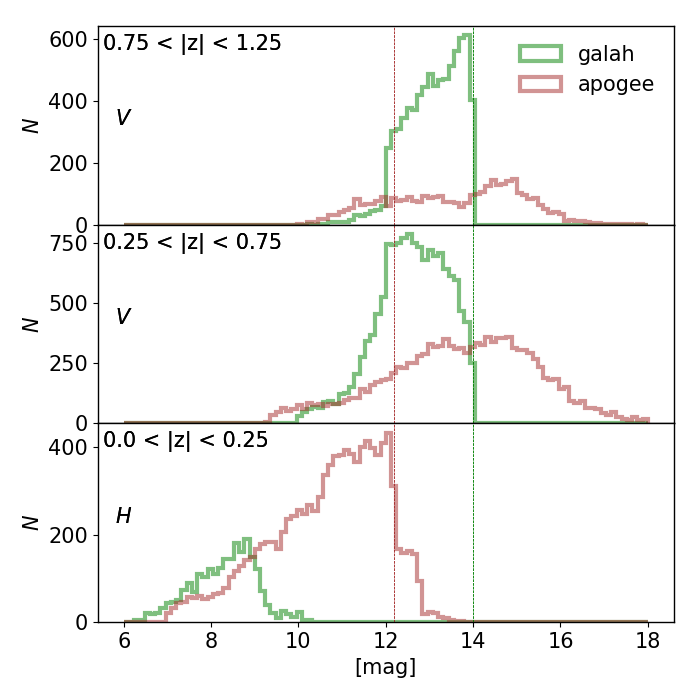}
\caption{Magnitude distribution for \glh{} and \apg{} Red Clump datasets shown for different slices in z (kpc). The position where the magnitude distribution falls sharply sets the maximum distance upto which the stars are unbiased in distance (pseudo volume complete). The magnitude limits are shown by dotted vertical lines. For the off-plane slices the completeness limit is $V=14$ and set by the \glh{} survey 
that dominates the number counts. For the mid-plane slice 
the limit is $H=12$ and set by \apg{} survey that  
dominates the number counts. 
\label{fig:GADR14_mag}}
\end{figure}
%============================

We now discuss the results of our kinematic modeling on the observed datasets and will compare this with selection function matched mock data generated with \glx{} as described in \autoref{sec:data}. Using Red Clump stars from \apg{} -DR12, B15 showed that after subtracting an axisymmetric model there remains a high residual in the \vlos{} field in the mid-plane ($|z|<0.25$ kpc). Their kinematical model assumed a flat rotation curve with $V_{\mathrm{circ}}=220$ \kms{} and $V_{\odot}=22.5$ \kms{} and the asymmetric drift was based on the Dehnen distribution \citep{1999AJ....118.1201D}. In \autoref{fig:power_spectrum_mid-plane} we consider again the B15 result and explore effects that can lead to enhanced residuals. In  \autoref{fig:power_spectrum_mid-plane}(a,b) we have reproduced their result by using the same model and data (\apg{} -DR12 RC sample) as them. 
A sharp peak of 10.4 \kms{} is obtained at a physical scale of about 2.5 kpc similar to B15. 

The location and the height of the peak is essentially unchanged when we include \apg{} -DR14 RC sample, the peak only becomes sharper (\autoref{fig:power_spectrum_mid-plane} c,d). 

Now, B15 used median statistics to compute the residual maps and power spectra. If the distribution of the residual velocity is a Gaussian then employing either mean or median statistics should not make much of a difference in the residual maps. However, if the distribution is asymmetric then it will. In the context of the Galaxy, we know that the $V_{\phi}$ distribution is asymmetric \citep{2014ApJ...793...51S}. Typically one defines a kinematic model and then computes the model parameters that maximize the likelihood of the model given the data. For such a best fit model, it is not clear as to which statistics (mean or median) will give lower values in velocity residual maps. 
In \autoref{fig:power_spectrum_mid-plane}(e,f) we find that choosing mean statistics lowers the power by 1.0 \kms{} for the B15 model. We have checked and found that for our best fit \textit{globalRz} model the results remain unchanged for either choice of statistic. So from now on for the rest of our analysis we adopt to use the mean statistics for computing the velocity residual maps.
Next, we consider the volume completeness of the data sample. \autoref{fig:GADR14_mag} shows the magnitude distribution of the \glh{} and \apg{} Red Clump stars (\galahapogee{} dataset) in $V$ and $H$ passbands. In the mid-plane region most of the data is from \apg{} and there is a sharp fall around $H=12$. Similarly, \glh{} contributes significantly to the off-plane slices and the distribution falls off around $V=14$, reflecting the survey selection function. This fall-off limit ($m_{\lambda, \rm{max}}$) is the faintest magnitude to which stars are observed completely (strictly speaking we mean pseudo-random-complete or unbiased in distance selection) and so we can also estimate the maximum distance this would correspond to by modifying equation \ref{dmod} as,
%============================
% EQUATION: dmod limit
%
\begin{equation}
d_{\mathrm{mod, max}} \leq m_{\lambda,\mathrm{max}}-M_{\lambda}- \sigma_{M_{\lambda}}-A_{\lambda} .
\end{equation}
%============================
Using magnitude limits for each slice, extinction factor $A_{\lambda}$ , absolute magnitude $M_{\lambda}$ and its dispersion $\sigma_{M_{\lambda}}$ from \autoref{tab:tab_extinct},  we find $d_{\mathrm{max}}=4$ kpc for the mid-plane and $d_{\mathrm{max}}=3.25$ kpc for the off-plane regions. These distance limits are also visible in the scatter plots of \autoref{fig:datacov_xy_zslices}. In \autoref{fig:power_spectrum_mid-plane}(g,h) we apply the $d<4$ kpc distance cut, which removes the high-residual pixels (beyond $x_{\rm hc} > 5$ kpc) however, there is no noticeable change in the power spectrum compared to \autoref{fig:power_spectrum_mid-plane}(g,h) as the amplitude is still at 9.3 \kms{}. However, as a precaution, we will continue with the distance limits for the rest of the figures. 

Finally, we replace the B15 model with our flexible axisymmetric model from \autoref{sec:kinematic_model} and this has the effect of further reducing the power to 7.1 \kms{} in \autoref{fig:power_spectrum_mid-plane}(i,j)}. In 
\autoref{fig:power_spectrum_mid-plane}(k,l) we  consider the residuals for the combined dataset \galahapogee{} to increase the sample size 
and get essentially the same power spectrum as in \autoref{fig:power_spectrum_mid-plane}(i,j) with lower amplitude of 6.3 \kms{}. 
A characteristic pattern of blue in first quadrant, red in second and yellow in third as seen in previous cases is also visible here. 
To conclude, we find that in the mid-plane after accounting for various systematics and a more flexible model the power amplitude can be reduced significantly, though interestingly it cannot be reduced to zero or to the level expected purely  due to noise (pink region).

%============================
% FIGURE: Data vs Galaxia (globalRz) for all slices.
\graphicspath{{figures/}} 
\begin{figure*}  
\includegraphics[width=2.0\columnwidth]{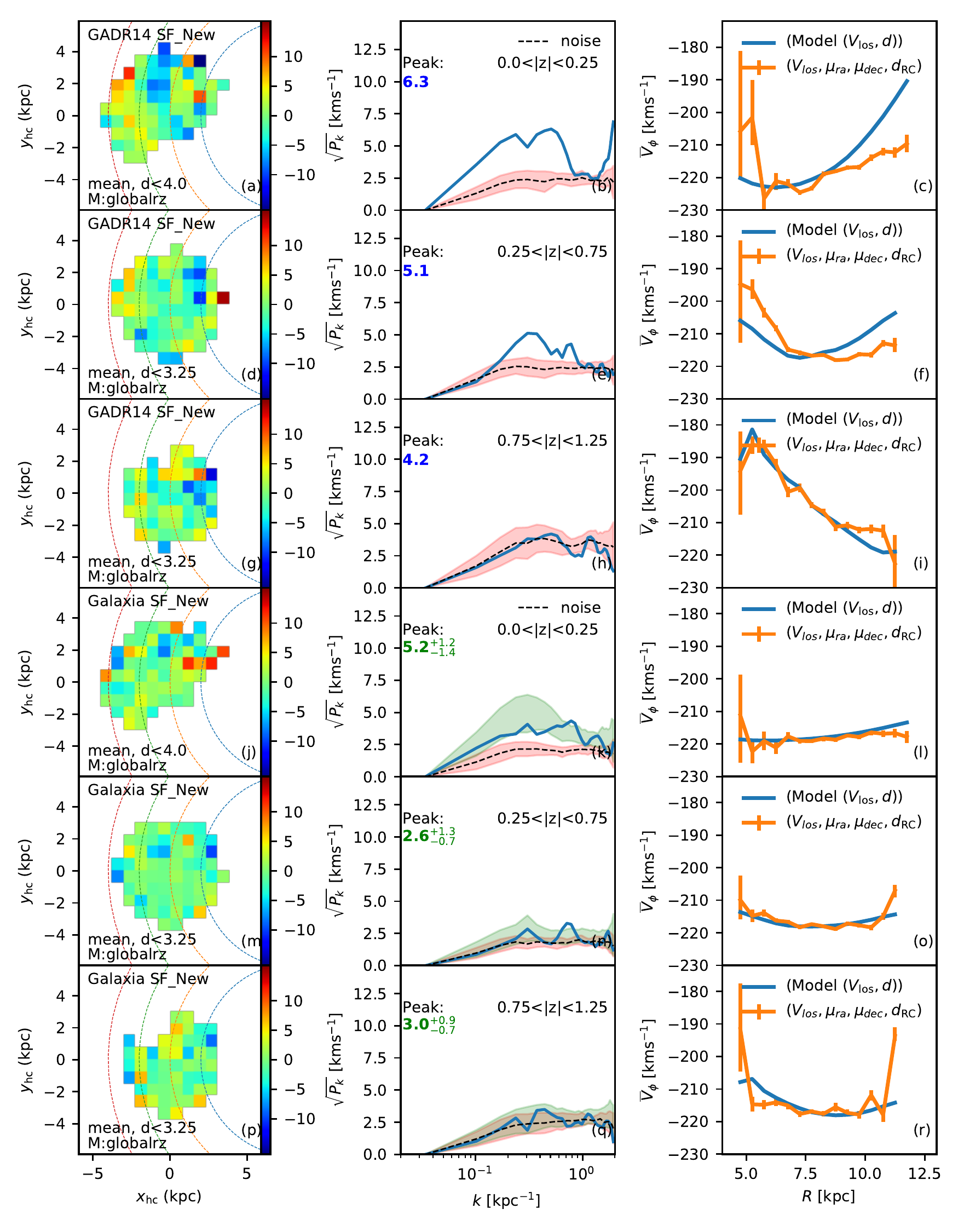}       
\caption{Residual $V_{\rm los}$ velocity maps, power spectrum and $\overline{V}_{\phi}$ profiles for observed and simulated data for different slices in $z$. The top three panels correspond to \galahapogee{} while the bottom three panels correspond to mock \glx{}. In each case, the power spectrum corresponding to the velocity map is shown in blue, the $1\sigma$ noise spread based on 20 random realisations in pink and median noise in dotted black. For \glx{} the green region represents the stochastic spread over 100 realisations, in power spectrum  with the same selection function as data. We find that except for the case of mid-plane the $\overline{V}_{\phi}$ profiles are captured well by the \textit{globalRz} models and the power spectrum of residual  $V_{\rm los}$ velocity approaches noise with amplitude $\approx 2$ \kms{}. \label{fig:power_spectrum_data_galaxia}}
\end{figure*}
%============================
%============================
% FIGURE: 
\graphicspath{{figures/}} 
\begin{figure*}  
\includegraphics[width=2.0\columnwidth]{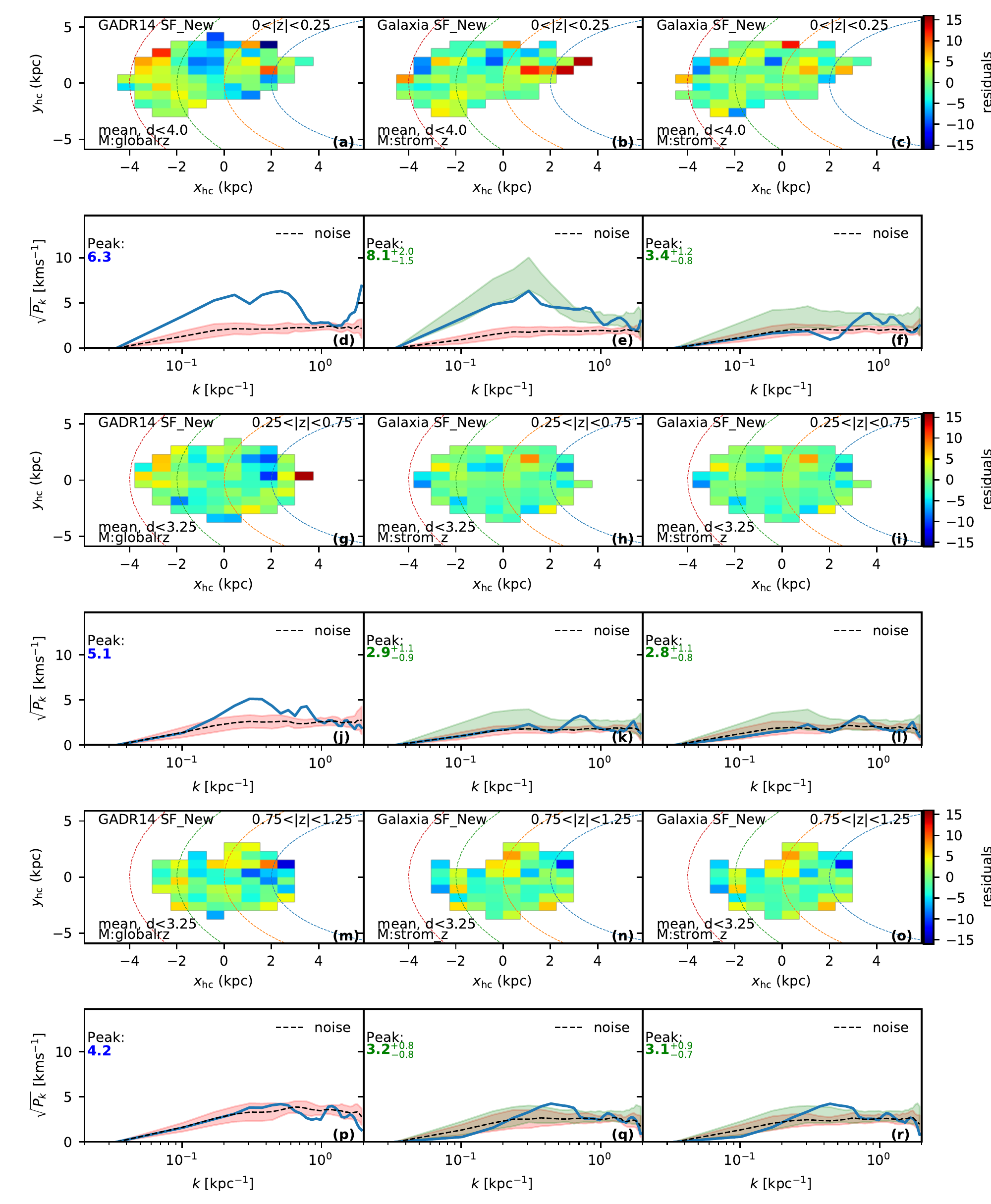}
\caption{{Residual velocity maps and power spectrum for observed and simulated data.
Left column shows results of the observed data. Middle column shows results for data simulated with \glx{}. Right column also shows results with \glx{} but when high mass stars with systematically underestimated distances are removed. First and second rows are for $|z/kpc|<0.25$, third and fourth are for $0.25<|z/kpc|<0.75$ and fifth and sixth are for $0.75<|z/kpc|<1.25$. For the observed data a 2d polynomial of degree 3 (nine coefficients) is employed to create the residual velocity map. For simulated data the kinematic model is based on the Str\"omberg equation and is known a priori.}\label{fig:power_spectrum_data_galaxia2}}
\end{figure*}
%============================
\subsection{Off-plane slices and comparison with \glx{}}
\label{sec:Data_vs_Galaxia}
We now also consider the off-plane ($|z|>0.25$ kpc) slices of data and also compare directly with mock realizations using \glx{}. Once again, we use the \galahapogee{} dataset and the flexible \textit{globalRz} model. In \autoref{fig:power_spectrum_data_galaxia}, we show the residual velocity maps, power spectra as well as the \mnvphi{} profile for each slice. 
To take the volume completeness of the sample into account, for the mid-plane slice we have restricted the data to $d<4.0$ kpc and for the off-plane slices to $d<3.25$ kpc.  

As mentioned already in \autoref{sec:mid-plane_data}, the peak  power in the mid-plane is around 6 \kms{} but moving away from the plane, the power drops (blue solid lines) and is only slightly higher than that expected from noise (dashed lines and the pink zone). Interestingly, the mock \glx{} samples also predict this trend of high power in the mid-plane but 
power that is lower and only slightly higher than noise elsewhere. 
Note, the predicted power spectrum has intrinsic stochasticity due to Poisson noise. So we generate 100 random realisations of the \glx{} samples and show the predicted 68\% confidence zone as the green shaded region. From these zones it is clear that, for \glx{} samples, the maximum power achieved in the mid-plane is $5.2^{+1.2}_{-1.4}$ \kms{}. For other slices, 
for \glx{} samples, the green and pink zones are almost 
on top of each other. 
However, the maximum power in observed data sets is higher by about 2 \kms{} as compared to \glx{} samples. 

We note that for the observed data and the $0.75<|z|<1.25$ slice the \mnvphi{} profile obtained using only line-of-sight motion traces well the \mnvphi{} profile obtained using both line-of-sight and proper motions. This suggests that, for this slice, there is minimal systematic error associated with distance, proper motion or line-of-sight velocities. However, for the other two slices which are closer to the plane 
we do see differences. The slice closest to the plane 
shows most pronounced deviations. The mock \glx{} samples 
also show similar behavior.
This is most likely due to systematic errors in distances as discussed in \autoref{sec:high_mass_stars}. If there are systematic errors 
with distances then its effect on the inferred 
\mnvphi{} profile will be different depending upon 
if we infer the profile based on line-of-sight velocities or both line-of-sight velocities and proper motions. 

The shape of the rotation profiles for the mock and observed data sets also show differences. For the mock data, the \mnvphi{} profile is predominantly flat across all the slices. In contrast, for the observed data a clear variation with $R$ is visible, and the variation becomes more pronounced as we move further away from the mid-plane. 
While our model is flexible enough to account for simple radial trends in rotation curves, this flexibility can over fit the data if the spatial coverage is not uniform. This is particularly a concern in the mid-plane where the coverage in the $(x,y)$ plane is not uniform, as there is a dearth of stars in the fourth quadrant.
This is because both APOGEEE and GALAH have not observed 
enough stars in the midplane and in the Southern Sky.  

Basically the constraints on \mnvphi{} for $R<8$ kpc come from data in the first and the fourth quadrant. 
As evidenced by the red and blue patches in \autoref{fig:vmap_hmag+rogue_allsky}b, the systematics in distances lead to incorrect values for the mean \vlos{} in the first and the fourth quadrant. If data from only one quadrant is available the model can adjust the value of $V_{\phi}$ for $R<8$ kpc to fit the \vlos{} in that quadrant perfectly, however this will not match the 
mean \vlos{} in the other quadrant. If the data from the other quadrant was also available the model would not have the freedom to do this, but in the absence of it the model over fits the data. 

The \glx{} samples are generated from a simulation for which the kinematics are known by design, so we can avoid over fitting a model which is similar to the input model. The input model has kinematics as a function of age, but since we do not have 
ages in the observed data, we approximate the kinematics by \textit{Strom\_z} model which is based on the Str\"omberg equation and described in \autoref{sec:kinematic_model}. In \autoref{fig:power_spectrum_data_galaxia2}, we employ this new fitting model \textit{Strom\_z} for \glx{} and compare its results with that of \textit{globalRz} model fitted to the observed data. Overall the trends in velocity maps and the power spectrum for the different slices are the same as in \autoref{fig:power_spectrum_data_galaxia}, i.e., high power in the mid-plane and negligible power away from the plane. The characteristic pattern of red in first quadrant and yellow in third as seen in observed data for the midplane slice is also reproduced in the midplane slice of the simulated data. For the \glx{} samples there is a slight increase in the power by about 3 \kms{} for the \textit{Strom\_z} as compared to \textit{globalRz}. This is not surprising, as the \textit{globalRz} model is more flexible and has more degrees of freedom than the \textit{Strom\_z} model. Moreover, in the plane \textit{globalRz} model can overfit the data due to incomplete coverage of the (x,y) plane. 

In \autoref{sec:high_mass_stars}, we showed that the presence of high mass RC stars can contaminate the kinematics in the mid-plane and can give rise to high \vlos{} residuals. \autoref{fig:power_spectrum_data_galaxia2}(c,f) shows that if we remove this population, by restricting stars to $d_{\rm RC}/d_{\rm true} > 0.8$, the excess power disappears. This suggests that the observed excess power is spurious and is due to contamination from high mass stars whose distances are underestimated. For the off-plane slices this additional cut makes no difference as the density of high mass RC is negligible for these slices.

%%%%%%%%%%%%%%%%%%%%%%%%%%%%%%%%%%%%%%%%%%%%%%%%%%
% Discussion
%%%%%%%%%%%%%%%%%%%%%%%%%%%%%%%%%%%%%%%%%%%%%%%%%%
\section{Discussion and Conclusions} \label{sec:discussion}
Over the past few years several surveys have hinted at non-axisymmetric motion in the disc of the Milky Way. \cite{2015ApJ...800...83B} used Red Clump stars from \apg{} to show velocity fluctuations of 11 \kms{} in the mid-plane region on scales of 2.5 kpc. In this paper we have made use of all the \apg{} Red Clump stars available up to date along with data from \glh{}. Our results do not dispute the presence of deviation from mean axisymmetric motion in the mid-plane of the Galaxy. However, simulations using \glx{} show that RC samples are likely to be contaminated by intrinscially brighter Red Clump stars, these stars are young and have high mass. Distance is underestimated for such stars. Being young, such stars lie preferentially closer to the midplane. This has the effect of contaminating the population at any given location with distant stars in that direction whose kinematics is different. This results in strange features when residual velocity maps are constructed in the $(x,y)$ plane. 

From \autoref{fig:power_spectrum_mid-plane}, we conclude that for the mid-plane slice the peak power $p_{\rm max}$
occurs at physical scales of $k^{-1} \approx 3$ kpc for the observed data, and is either 9.3 \kms, using the original \textit{Bovy1} model, or 6.3 \kms, using the more flexible axisymmetric model \textit{globalRz}. On the other hand, the simulations from \glx{} in Figures \ref{fig:power_spectrum_data_galaxia} and \ref{fig:power_spectrum_data_galaxia2} show that the peak power is $8.1^{+2.0}_{-1.5}$ \kms  using the \textit{Strom\_z} model or $5.2^{+1.2}_{-1.4}$ \kms  with the flexible \textit{globalRz} model. 
The peak in the power spectrum is also at the same physical scale of 3 kpc for both the observed sample and \glx{} sample. We have also demonstrated that the power in \glx{} is due to contamination from young high mass Red Clump stars, as the sample with $d_{RC}/d_{\rm True}>0.8$ does not show excess power.  
So we do expect the high mass stars to  contribute to the power in the observed data, but how much is the contribution from real streaming motion is not obvious at this stage. 
The streaming and spurious perturbations in the velocity field could be correlated or uncorrelated. For the first case the streaming perturbations will add on to spurious perturbations and will enhance the power linearly. This would mean that the real streaming motion (observed peak power minus the average predicted peak power by \glx{} is less than 1.2 \kms{}, adopting either \textit{StromR\_z} or \textit{globalRz} as the reference model. Note, the observed fluctuations using  the \textit{Bovy1} model are best 
compared with \glx{} predictions using the \textit{Strom\_z} model, 
as both models are inflexible models.
If instead they are uncorrelated, we would expect the contributions to be added quadratically (given that power is physically a measure of dispersion), leading to an estimate of 4.6 \kms{} using \textit{StromR\_z} and 3.6 \kms{} 
using \textit{globalRz}. 

In the mid-plane using the flexible $globalRz$ model we have been able to reduce the power from 9.3 to 6.3 \kms. The red pattern in the first quadrant and the yellow in the third are subdued. 
However, the blue pattern in second quadrant still exists, 
which could be due to a real feature in the data. 

For slices away from the plane, $0.25<|z/ kpc|<0.75$ and 
$0.75<|z/ kpc|<1.25$, we find that for the observed data the power decreases with height above the plane and is no more than 5.1 \kms. 
This rules out large non-axisymmetric streaming  motion extending beyond the $|z| > 0.25$ kpc. The \glx{} samples also 
predict very little power ($3$ \kms) for slices away from the plane. However, the power in the observed data is higher than 
that predicted by \glx{} by about 2 \kms. So, small 
streaming motion is not ruled out. Assuming streaming motion to be uncorrelated with other effects, we estimate the power to be less than 4.4 \kms for $0.25<|z/ kpc|<0.75$ and less than 2.9 \kms for 
$0.75<|z/ kpc|<1.25$.

If the excess power in the observed data is real and not 
an artefact of high mass clump stars, then it is interesting 
to consider the cause behind the decrease of power with height. This could be indicative of the fact that it is much easier to 
excite streaming motion in young dynamically cold populations 
than old dynamically hot populations.

We note that the analysis presented here has limitations when applied to data away from the mid-plane. The average 
age of stars increases with height above the plane due to the age scale height relation in the Galaxy.
The mean azimuthal motion depends upon age and hence is also a function of $|z|$. Now, if a slice in  $|z|$ is not sampled 
uniformly in the $(x,y,z)$ space, the mean residual motion will show large variance just due to incomplete sampling.
It is quite common for spectroscopic surveys to have such incomplete sampling at high $|z|$, as they observe in small patches across the sky. In such cases, one should always compare the power spectrum of observed data with selection 
function matched mock data which will correctly capture the power due to incomplete sampling. 

Finally, \cite{2015ApJ...800...83B}, using their axisymmetric model, obtained a power excess in the mid-plane region, of $\approx12$ km $^{-1}$ and strongly suggested that the LSR itself is streaming at this velocity. They add this excess to the \cite{2010MNRAS.403.1829S} value for the Sun's peculiar motion to give the new $V_{\odot}\approx 12.1 + 12.0=24.1$ \kms. Following our analysis, we suggest that the adjustment to $V_{\odot}$ should be no more than 4.2 km $s^{-1}$, provided the excess power in the residual velocity field is not due to high-mass Red Clump stars. Interestingly, \cite{2018MNRAS.tmp.2508K} using \gaia{} DR1 Cepheids also obtain $V_{\odot} = 12.5 \pm 0.8 $ \kms{} i.e., consistent with \cite{2010MNRAS.403.1829S}, although they do not assert it to be conclusive given the small size of their sample.

We find that the spectro-photometric RC selection criterion given by \cite{2014ApJ...790..127B} is quite efficient at isolating the  RC stars. Based on \glx{} simualtions, the criterion can isolate RC stars with a purity of 98\%. We further refined the criteria and made it purely based on spectroscopic parameters. However, we find that such 
selection criteria have a strong age bias, Red Clump stars below 2 Gyrs are significantly underrepresented.

Looking further to the future, \gaia{} can resolve some of the questions raised by our analysis. First, with accurate parallaxes from \gaia{}, we can confirm if the \apg{} Red Clump catalog contains high mass stars with underestimated distances. If so, then does removing this population get rid of the excess power 
in the residual velocity map? Moreover, with proper motion we can construct and study velocity maps of $V_{\phi}, V_{R}$ and $V_z$ separately instead of just \vlos{}. We can also make use of all type of stars and not just the Red Clump.

%%%%%%%%%%%%%%%%%%%%%

\section*{Acknowledgements}
The GALAH survey is based on observations made at the Australian Astronomical Observatory, under programmes A/2013B/13, A/2014A/25, A/2015A/19, A/2017A/18. We acknowledge the traditional owners of the land on which the AAT stands, the Gamilaraay people, and pay our respects to elders past and present. Parts of this research were conducted by the Australian Research Council Centre of Excellence for All Sky Astrophysics in 3 Dimensions (ASTRO 3D), through project number CE170100013. S.S. is funded by University of Sydney Senior Fellowship made possible by the office of the Deputy Vice Chancellor of Research, and partial funding from Bland-Hawthorn's Laureate Fellowship from the Australian Research Council. DMN was supported by the Allan C. and Dorothy H. Davis Fellowship.

This research has made use of Astropy13, a community-developed core Python package for Astronomy (Astropy Collaboration et al., 2013). This research has made use of NumPy (Walt et al., 2011), SciPy, and MatPlotLib (Hunter, 2007).

Funding for the Sloan Digital Sky Survey IV has been provided by the Alfred P. Sloan Foundation, the U.S. Department of Energy Office of Science, and the Participating Institutions. SDSS-IV acknowledges
support and resources from the Center for High-Performance Computing at
the University of Utah. The SDSS web site is www.sdss.org.

SDSS-IV is managed by the Astrophysical Research Consortium for the 
Participating Institutions of the SDSS Collaboration including the 
Brazilian Participation Group, the Carnegie Institution for Science, 
Carnegie Mellon University, the Chilean Participation Group, the French Participation Group, Harvard-Smithsonian Center for Astrophysics, 
Instituto de Astrof\'isica de Canarias, The Johns Hopkins University, 
Kavli Institute for the Physics and Mathematics of the Universe (IPMU) / 
University of Tokyo, Lawrence Berkeley National Laboratory, 
Leibniz Institut f\"ur Astrophysik Potsdam (AIP),  
Max-Planck-Institut f\"ur Astronomie (MPIA Heidelberg), 
Max-Planck-Institut f\"ur Astrophysik (MPA Garching), 
Max-Planck-Institut f\"ur Extraterrestrische Physik (MPE), 
National Astronomical Observatories of China, New Mexico State University, 
New York University, University of Notre Dame, 
Observat\'ario Nacional / MCTI, The Ohio State University, 
Pennsylvania State University, Shanghai Astronomical Observatory, 
United Kingdom Participation Group,
Universidad Nacional Aut\'onoma de M\'exico, University of Arizona, 
University of Colorado Boulder, University of Oxford, University of Portsmouth, 
University of Utah, University of Virginia, University of Washington, University of Wisconsin, 
Vanderbilt University, and Yale University.

%%%%%%%%%%%%%%%%%%%%%%%%%%%%%%%%%%%%%%%%%%%%%%%%%%

%%%%%%%%%%%%%%%%%%%% REFERENCES %%%%%%%%%%%%%%%%%%

% The best way to enter references is to use BibTeX:

\bibliographystyle{mnras}
\bibliography{biblio.bib} % if your bibtex file is called example.bib

%%%%%%%%%%%%%%%%% APPENDICES %%%%%%%%%%%%%%%%%%%%%

\appendix

\section{Red Clump calibration and selection}
\label{app:red_clump_sel}  

%============================
% FIGURE: log-Teff for dwarfs/giants and Red Clump
\graphicspath{{figures/}}
\begin{figure}
\centering
\includegraphics[width=1.0\columnwidth]{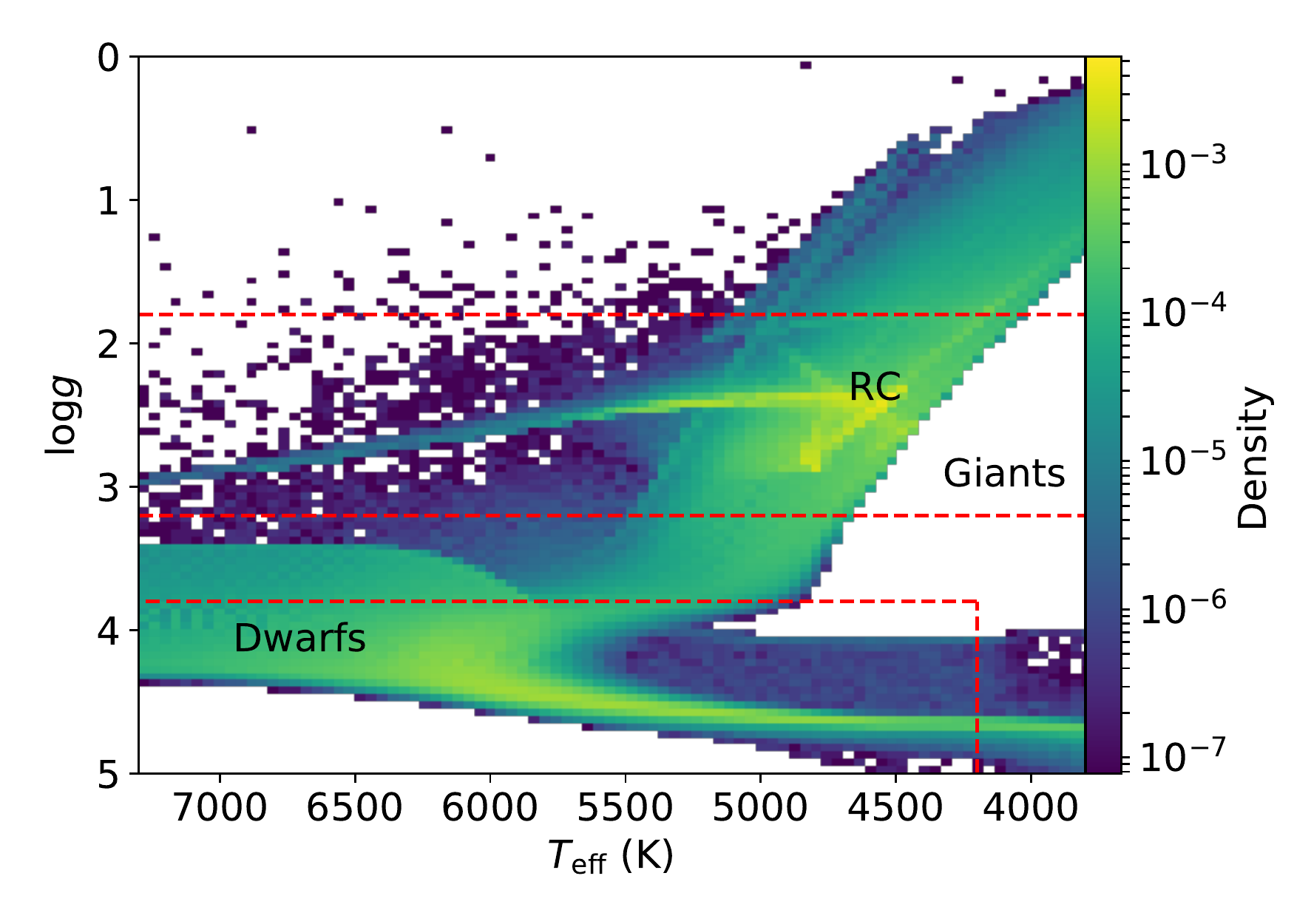}          
\caption{Spectroscopic HR diagram of the \glx{} $J<15$ all-sky sample used to derive \clr{} as a function of \teff{} and \logg{}. Red dashed lines mark the approximate boundary between Dwarfs and Giants and the typical location of Red Clump (RC) is indicated.{\label{fig:logg_teff_pop}}}
\end{figure}
 %============================
Following from \autoref{sec:red_clump_sel_brief}, here we describe details of our Red Clump selection and calibration. A crude sample of RC stars can be selected based on cuts in surface gravity \logg{} and dereddened color \clr, for example, \cite{2013MNRAS.436..101W} used the simple cuts of $1.8 \leq \logg \leq 3.0$ and $0.55<(J-K)_{0}<0.8$ on \rave{} data. However, this was estimated to be contaminated by about 30-60 \% of non-RC stars, including the secondary Red Clump \citep{1999MNRAS.308..818G}, and the red giant branch bump, which is a metallicity-dependent localized excess in the luminosity function of first-ascent red giant branch stars \citep{1997MNRAS.285..593C,2013ApJ...766...77N}.

In the APOGEE-RC catalog \cite{2014ApJ...790..127B} use PARSEC isochrones \citep{2012MNRAS.427..127B} and asteroseismic constraints to improve the sample purity, resulting in the following comprehensive selection scheme
%--
\begin{eqnarray}
1.8 \leq \logg \leq 0.0018\, \mathrm{dex}\, \mathrm{K}^{-1}\,\,\Big(\teff-\teff^{\mathrm{ref}}(\feh)\Big)+2.5\ , \label{logg_cut}\\
Z  >  1.21 [(J-K)_{0} -0.05]^{9} + 0.0011 \label{Z_Cut1},\\
Z  <  {\rm Min}\left(2.58 [(J-K)_{0} -0.40]^{3} + 0.0034, 0.06\right) \label{Z_Cut2},\\
0.5  < (J-K)_{0} <0.8 \label{clr_cut},
\end{eqnarray}
%--
where,
%--
\begin{equation}
\teff^{\mathrm{ref}}(\feh) = -382.5\,\mathrm{K} \, \mathrm{dex}^{-1}\, \feh+4607\,\mathrm{K}\ .
\end{equation} and Z is the PARSEC isochrone metallicity.
%--
However, this requires de-reddened \jkzero{} color and to get them extinction is required. In the \apg{} Red Clump catalog by \cite{2014ApJ...790..127B} 
extinction was estimated using the Rayleigh Jeans Color Excess method \citep[RJCE;][]{2011ApJ...739...25M} which requires photometry in [4.5$\mu$] band. 
Extinction estimates based purely on photometry are useful but have inaccuracies associated with them. To overcome this, we use pure Red Clump stars from \glx{} to derive empirical relations expressing \jkzero{} in terms of spectroscopic parameters 
[Fe/H] and \teff{}. 
Such relations have previously been derived for K-type dwarf stars by \cite{2010A&A...512A..54C}, where one fits for a function of the form
%--
\begin{equation}
\label{calib_eqn}
5040\, \mathrm{K}/T_{\rm eff} = a_{0} + a_{1}X + a_{2}X^{2} +a_{3}XY + a_{4}Y + a_{5}Y^{2} , 
\end{equation}
%--
where X = \clr{}, Y = [Fe/H] and $(a_{0}...a_{5})$ are the fit coefficients. While this is a valid function to use, it is not analytically invertible to derive \clr, unless the dependence on [Fe/H] can be neglected in which case equation \ref{calib_eqn} can be easily inverted to give\footnote{For completion we also perform the fitting using equation \ref{calib_eqn} with and without the [Fe/H] term and found that the derived temperature had residuals below 20 K for Red Clump and Giants, though Dwarfs had higher ($\approx$50 K) residuals without the [Fe/H] term. We provide these results in \autoref{tab:calib_table_indirect} but do not use it for our analysis in this paper.}
%--
\begin{equation}
\label{teff2clr}
(J-K)_{0} \sim \frac{1}{2a_{2}} [-a1 + \sqrt{a_{1}^{2} - 4a_{2} \large(a_{0} - \frac{5040\, \mathrm{K}}{T_{\rm eff}} \large) ]} .
\end{equation}
%--
We show below that the dependence on [Fe/H] is weak but not negligible. So we alter equation \ref{calib_eqn} to fit directly for \clr{} as
%--
\begin{equation}
\label{direct_calib_eqn}
(J-K)_{0} = a_{0} + a_{1}X + a_{2}X^{2} +a_{3}XY + a_{4}Y + a_{5}Y^{2} ,
\end{equation}
%--
where $X =$[Fe/H] and $Y=5040\, \mathrm{K}/T_{\rm eff}$.

To derive the coefficients, we use data simulated by the code \glx{}, which allows us to obtain relations valid for majority of the stars that we observe. More specifically, we generate an all-sky catalogue with $J<15$, where the stellar parameters are generated using PARSEC\footnote{The isochrones were downloaded from \href{http://stev.oapd.inaf.it/cmd}{http://stev.oapd.inaf.it/cmd }} isochrones 
\citep{2012MNRAS.427..127B,2014MNRAS.444.2525C,2015MNRAS.452.1068C,2014MNRAS.445.4287T}, and choose the 2MASSWISE photometric system. From this we select three populations using boundaries in \logg, namely Dwarfs ($\logg \geq 3.8$), Giants ($\logg \geq 3.2$), and Red Clump stars ($1.8 \leq \logg \leq 3.0$). \autoref{fig:logg_teff_pop} marks the approximate boundaries between the three populations in the spectroscopic HR  diagram. 
For a given age and metallicity of a star, stellar models can predict the initial mass required to reach the tip of the giant branch, and so for Red Clump stars the initial mass must exceed this threshold tipping mass (i.e $>M_{\rm RGB, tip}$). We make this additional cut to identify the real Red Clump stars in \glx{}. We also exclude M-dwarfs from our analysis by applying a temperature cut of $4200<T_{\rm eff} (K)<8000$, as the $(J-K)$ color is not 
a good indicator of temperature for them. 

%============================
% FIGURE: calibration fits
\graphicspath{{figures/}} 
\begin{figure*}
\centering    
\includegraphics[width=2.0\columnwidth]{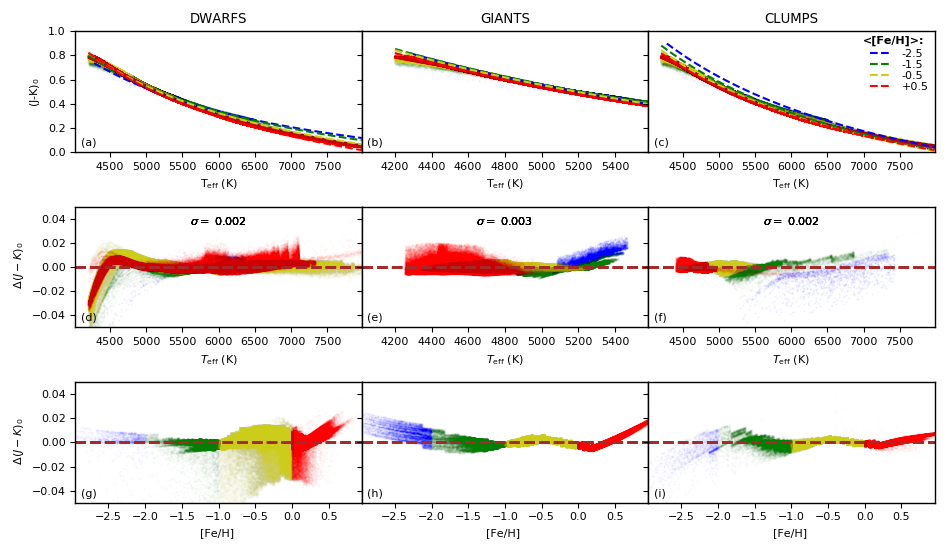}
\caption{{Empirical (\clr{} |\teff{}, \logg{}) calibration using \glx{} all-sky sample based on equation \ref{direct_calib_eqn}: Panels (a,b,c) show the best-fit (dotted curves) for Dwarfs, Giants and Red Clump stars and the color-coding is the mean [Fe/H]. Panels (d,e,f) and (g,h,i )show the residuals against temperature \teff{} and [Fe/H] respectively. While for Dwarfs the derived relations fit well at low metallicity and high temperatures, for the Giants and Red Clump, residuals are low everywhere except for at [Fe/H]<-2.}{\label{fig:calib_plts}}}
\end{figure*}
%============================ 
 
The resulting best-fit coefficients $(a_{0}...a_{5})$ for each population are listed in \autoref{tab:calib_table}, using which we derive $(J-K)_{0}$. 
Figure \ref{fig:calib_plts} shows the predicted $(J-K)_{0}$ and residuals as a function of \teff{} and [Fe/H]. The best-fit curves trace the color well and the residuals for all three populations are below 0.003 mag. As mentioned earlier, weak metallicity dependence is visible. For the Red Clump and giants, the residuals  show  very little variation with temperature (\autoref{fig:calib_plts}e and f), but with metallicity (\autoref{fig:calib_plts}h and i) a systematic effect can be seen for [Fe/H]$<-2$.  In comparison for dwarfs  higher metallicities and lower temperatures have high residuals (\autoref{fig:calib_plts}d and g).

\begin{figure*}%!t
\centering
\includegraphics[width=1.5\columnwidth]{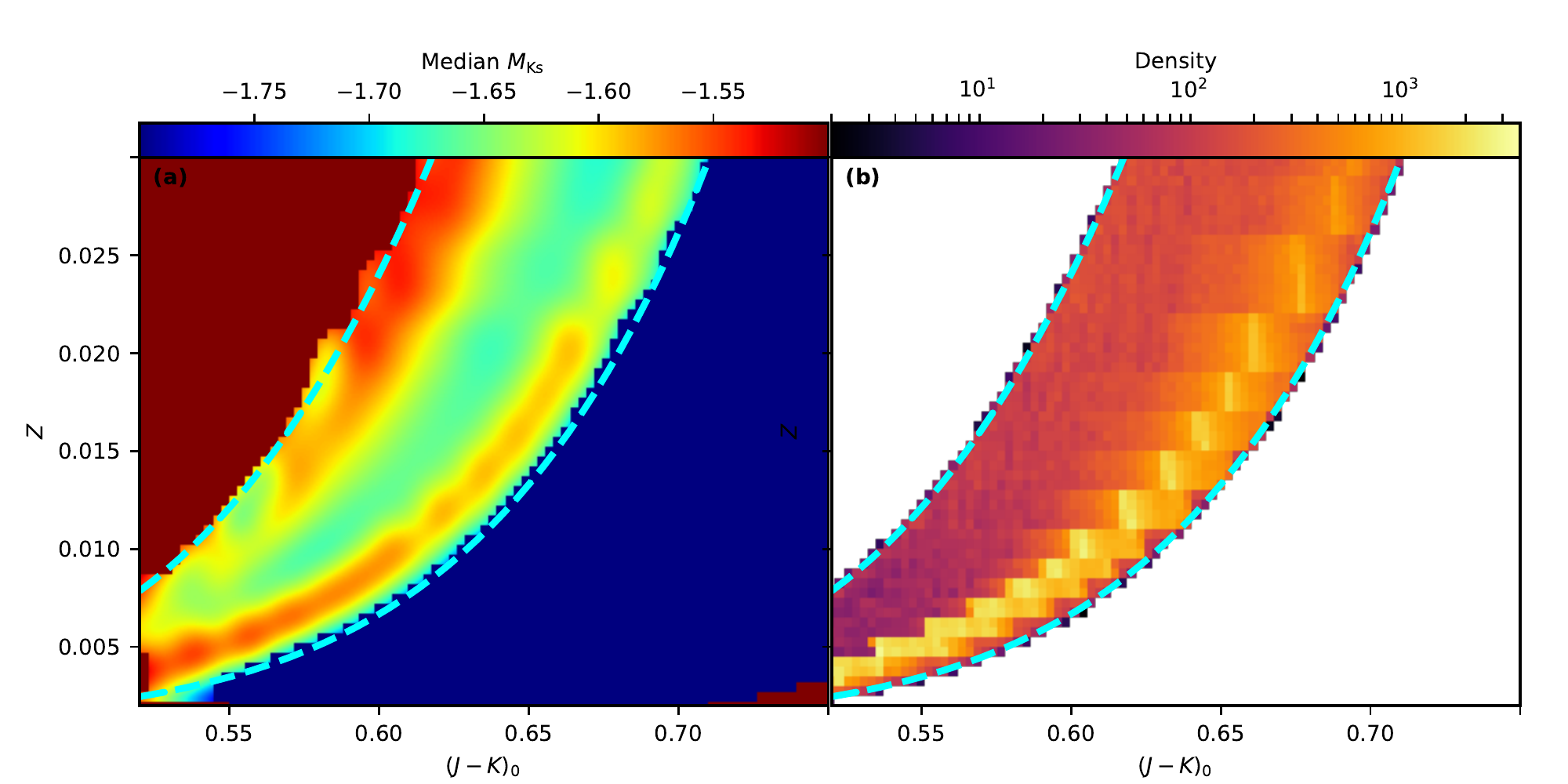}
\caption{The all-sky \glx{} Red Clump sample selected using \teff{} and \logg{} cuts based on equations \ref{logg_cut} and \ref{clr_cut}, and with additional cuts (cyan dotted line) based on equations \ref{Z_Cut1} and \ref{Z_Cut2} using Color-Temperature-metallicity calibration necessary to remove contamination from non RC stars. In the final selected sample, the median $M_{\rm k}$ lies in a narrow band around -1.60 (Panel a) and most stars are concentrated around this value (Panel b).\label{fig:rc_selection_met_clr}}
\centering
\includegraphics[width=1.5\columnwidth]{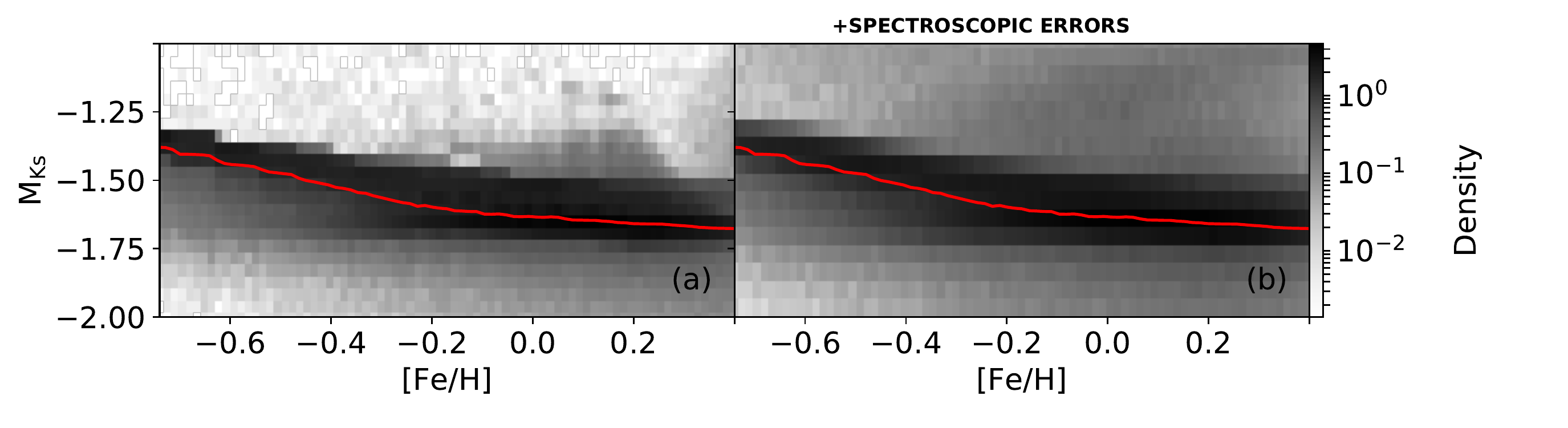}
\caption{{$M_{\rm K}$ - [Fe/H] distribution for the \glx{} Red Clump sample.
(a) This is the case without any errors in spectroscopic parameters. 
The sample has a tight distribution with dispersion in estimated distance modulus of $\sigma_{\rm dmod}=0.1$; the running average (solid red curve) can thus be  used to derive absolute magnitude $M_{\rm K}$ from spectroscopy. (b) 
This is the case with spectroscopic errors  of $(\sigma_{\log T_{\rm eff}}, \sigma_{\rm [Fe/H]}, \sigma_{\log g})= (0.011,0.05,0.1)$ dex. Here dispersion 
in distance modulus increases to $\sigma_{M_{\rm k}}=0.18$. The red curve is same  as in (a).}{\label{fig:rc_selection}}}
\end{figure*}

%====================
% TABLE: Direct Calibration
\begin{table}
\centering
\caption{Best-fit coefficients for equation \ref{direct_calib_eqn} used to derive \jkzero{} for the three populations: Dwarfs, Giants and Red Clump. The fitting is carried out over the temperature range 4200$<$\teff$<$8000.\label{tab:calib_table}}
\begin{tabular}{l|l|l|l|l|l|l}%[!hb]%[!htbp]
\hline
\hline
Population & $a_{0}$ & $a_{1}$ & $a_{2}$ & $a_{3}$ & $a_{4}$ & $a_{5}$ \\
\hline
Dwarfs & -0.637 & -0.107 & -0.007 & 0.093 & 0.915 & 0.251 \\	
Giants & -0.957 & 0.000 & -0.006 & -0.020 & 1.489 & 0.002 \\
Red Clump & -0.800 & 0.046 & 0.008 & -0.060 & 1.199 & 0.132 \\
\hline
\end{tabular}
\end{table}
%============================ 
Finally using these derived colors we can now use equations \ref{logg_cut}-\ref{clr_cut} to produce a sample of Red Clump stars from our mock $J<15$ \glx{} catalogue. Here and throughout the paper for the purpose of selection function we make use of the $(J-K)_{0}$ relation corresponding to the Red Clump stars. \autoref{fig:rc_selection_met_clr} shows the Red Clump selection in metallicity-color space and illustrates the effect of applying additional cuts from equations \ref{Z_Cut1} and \ref{Z_Cut2} (using color-temperature-metallicity selection) in order to remove contamination from non-RC stars. It is clear that the final selection has a very narrow range in the median $M_{\rm k}$ and lies around -1.60.

With the RC sample selected, in \autoref{fig:rc_selection} we plot the $M_{\rm K}$ against [Fe/H] and a running median curve (shown in red) that can be used to approximate Red Clump magnitude from metallicity. The dispersion in estimated distance modulus ($\sigma_{\rm dmod}$) increases from 0.1 to 0.16 by adding spectroscopic errors, however, if the uncertainty in temperature is a factor of two lower this lower $\sigma_{\rm dmod}$ to 0.12. For the GALAH data we can get such precision for good signal to noise data \citep{2018MNRAS.476.5216D,2018MNRAS.473.2004S}.

%====================
% TABLE: Magnitude/A(i)/sigma(i)
\begin{table}
\centering
\caption{Median absolute magnitude $M_{\mathrm{RC}}$, and dispersion in absolute magnitude $\sigma_{M_{\mathrm{RC}}}$ for Red Clump stars selected from \glx{} using the scheme in \autoref{app:red_clump_sel}. We have tabulated the values for a few common passbands only for a comparison with literature. Also listed are the extinction factors ($f_{\lambda}$) for the four passbands and these are taken from \citep{1998ApJ...500..525S}.\label{tab:tab_extinct}}
\begin{tabular}{llll}
\hline
Passband ($\lambda$) & $M_{\mathrm{RC}}$ & $\sigma_{M_{\mathrm{RC}}}$ & $f_{\lambda} = \frac{A_{\lambda}}{E(B-V)}$ \\
\hline
$J$      & $-0.98  $ & $0.11  $ &  0.902 \\	
$H$      & $-1.52  $ & $0.12  $ &  0.576 \\
$K$      & $-1.60  $ & $0.13  $ &  0.367 \\
$V_{JK}$ & $+0.75 $ & $0.15 $ & 3.240 \\
\hline
\end{tabular}
\end{table}
%====================

For some simple calculations it is useful to know the typical absolute magnitude 
of Red Clump stars in different photometric bands, e.g., to estimate the 
volume completeness of various surveys. 
Hence, in \autoref{tab:tab_extinct} we list the median absolute magnitude and dispersion based on 68\% confidence region for the $J,H,K$ and $V_{JK}$ pass bands. Here,  
\begin{equation}
V_{JK}=Ks + 2.0 (J-Ks+0.14)+0.382\exp[(J-Ks-0.2)/0.50]
\end{equation}
is the Johnson $V$ band magnitude computed using \twomass{} magnitudes \citep{2018MNRAS.473.2004S}.
Our derived values are in good agreement with literature \citep{2016ARA&A..54...95G}.
%============================ 
% FIGURE: rc selection 
\graphicspath{{figures/}} 
%============================ 

%====================
% TABLE: Mks-[Fe/H] curve tabulated
\begin{table*}
\centering
\caption{Tabulated values for mean absolute magnitude $M_{\rm{Ks}}$ as function of [Fe/H] as derived with \glx{} (red line in \autoref{fig:rc_selection}a). Distances to Red Clump stars are estimated by linear interpolating over these values. \label{tab:mks_feh}}
\begin{tabular}{l|l|l|l|l|l|l|l|l|l|l|l|l|l}
\hline
[Fe/H]  (dex) & -0.8   & -0.7   & -0.6   & -0.5   & -0.4   & -0.3   & -0.2   & -0.1   & 0.0    & 0.1    & 0.2    & 0.3    & 0.4 \\
\hline
$M_{\rm{Ks}}$ & -1.390 & -1.405 & -1.442 & -1.475 & -1.520 & -1.564 & -1.598 & -1.622 & -1.633 & -1.646 & -1.659 & -1.666 & -1.676 \\
\hline
\end{tabular}
\end{table*}
%====================

%====================
% TABLE: Calibration
\begin{table*}
\centering
\caption{Best-fit coefficients for equation \ref{calib_eqn} for the three populations: Dwarfs, Giants and Red Clump stars. Equation \ref{calib_eqn} derives 5040 K/\teff{} which can be analyticaly be inverted to derive \jkzero{} if we neglect the [Fe/H] term. This alters the coeffecients slightly and so we also list that case. Note: This table is only provided for completion and we do not use it for our analysis in this paper.\label{tab:calib_table_indirect}}
\begin{tabular}{l|l|l|l|l|l|l|l|l}
\hline
Population & [Fe/H] used? & $a_{0}$ & $a_{1}$ & $a_{2}$ & $a_{3}$ & $a_{4}$ & $a_{5}$ & No. of stars\\
\hline
Dwarfs & yes & 0.5985 & 0.8148 & -0.104 & -0.053 & 0.0382 & 0.0045 & 212120 \\
Dwarfs & no & 0.6045 & 0.7700 & -0.051 & - & - & - & 212120 \\
Red Clump & yes & 0.6511 & 0.6410 & 0.0298 & -7e-05 & 0.0116 & -0.002 & 141666 \\
Red Clump & no & 0.5701 & 0.8421 & -0.083 & - & - & - & 141666 \\
Giants & yes & 0.6447 & 0.6651 & 0.0010 & 0.0044 & 0.0113 & 0.0042 & 135891 \\
Giants & no & 0.5458 & 0.9260 & -0.171 & - & - & - & 135891 \\
\hline
\end{tabular}
\end{table*}
%====================

\newpage
\section{Phase-Space transformation equations}
\label{app:trans_eqns}
For our main analysis we fit a model for the mean $V_{\phi,GC}$ to the \vlos{} data. For this we require the following transformation from Galactocentric to heliocentric coordinates:
\begin{equation}
(l,z,R,V_{\phi},V_{z},V_{R})_{\mathrm{GC}} \rightarrow (l,b,d,V_{l},V_{b},V_{\mathrm{los}})_{\mathrm{HC}}.
\end{equation}
This is achieved in the sequence,
\begin{itemize}
\item $(x,y,z)_{\mathrm{GC}} = \textbf{\textit{lzR2xyz}}(l,z,R)_{\mathrm{GC}}$,
\item $(V_{x},V_{y},V_{z})_{\mathrm{GC}} = \textbf{\textit{VlzR2xyz}}(V_{\phi},V_{z},V_{R})_{\mathrm{GC}}$,
\item $(V_{x},V_{y},V_{z})_{\mathrm{HC}} = (V_{x},V_{y},V_{z})_{\mathrm{GC}}- (U,\Theta,W)_{\odot}$,
\item $(x,y,z)_{\mathrm{HC}} = (x,y,z)_{\mathrm{GC}}-(x,y,z)_{\odot}$,
\item $(V_{l},V_{b},V_{\mathrm{los}})_{\mathrm{HC}} = \textbf{\textit{Vxyz2lbr}}(x,y,z,V_{x},V_{y},V_{z})_{\mathrm{HC}}$, 
\end{itemize}
where following \cite{2010MNRAS.403.1829S} we adopt $(U,V)_{\odot}=(11.1,7.25)$ km s$^{-1}$, and the azimuthal component $\Theta_{\odot} =242.0$ km s$^{-1}$ for data (239.08 for \glx{}). The $\Theta_{\odot}$ for data is estimated as $\Omega_{\odot}R_{\odot}$, with $R_{\odot}=8$ kpc and  $\Omega_{\odot}=30.24$ km s$^{-1}$kpc$^{-1}$ as set by the proper motion of Sgr A* \citep{2004ApJ...616..872R}. The transformation matrices (in bold) are defined in \autoref{tab:transfm_eqn}.\\

\begin{table}
\centering
\caption{Coordinate Transformation matrices. \label{tab:transfm_eqn}}
\begin{tabular}{c|c|c}
%\toprule
\textbf{\textit{lzR2xyz}} & \textbf{\textit{VlzR2xyz}} & \textbf{\textit{Vxyz2lbr}}\\
\hline
$\begin{bmatrix}  x  \\ y \\ z \end{bmatrix} = \begin{bmatrix} R cos(l)\\ R sin(l) \\ z \end{bmatrix}$; &
$\begin{bmatrix}  V_{x} \\  V_{y} \\  V_{z} \end{bmatrix} = \begin{bmatrix}  V_{\phi} & V_{z} &  V_{R}  \end{bmatrix}
\begin{bmatrix}  -sin(l)     &  cos(l)   & 0  \\  0 & 0 & 1 \\ cos(l) & sin(l) & 0 \end{bmatrix}$; &
$\begin{bmatrix}  V_{l} \\  V_{b} \\  V_{r} \end{bmatrix} = \begin{bmatrix}  V_{x} & V_{y} &  V_{z}  \end{bmatrix}
\begin{bmatrix}  -y/rc     & -zx/rc        & x  \\  -x/rc & -zy/rc & y \\ 0  & rc & z \end{bmatrix}$\\
\hline
& & $rc = \sqrt[]{x^{2} + y^{2}}$, $V_{r} = V_{\rm los}$\\ 
\hline 
\end{tabular}
\end{table}

On the other hand, to obtain the `true' rotation profiles, we first convert the longitudinal and latitudinal proper motions to heliocentric velocities:
\begin{eqnarray}
V_{l} = \mu_{l} \times d \times  4.74 \times 10^{3}   \\
V_{b} = \mu_{b} \times d \times  4.74 \times 10^{3}   
\end{eqnarray} and then combined with \vlos{} use the sequence above in reverse order to obtain \vphi{}.

%%%%%%%%%%%%%%%%%%%%%%%%%%%%%%%%%%%%%%%%%%%%%%%%%%

%%%%%%%%%%%%%%%%%%%%%%%%%%%%%%%%%%%%%%%%%%%%%%%%%%

% Don't change these lines
\bsp	% typesetting comment
\label{lastpage}
\end{document}